\documentclass[aps,prb,twocolumn,amssymb,amsmath,showpacs]{revtex4-2}

\makeatletter
  \long\def\pprintMaketitle{\clearpage
  \iflongmktitle\if@twocolumn\let\columnwidth=\textwidth\fi\fi
  \resetTitleCounters
  \def\baselinestretch{1}%
  \printFirstPageNotes
  \begin{center}%
 \thispagestyle{pprintTitle}%
   \def\baselinestretch{1}%
    \Large\@title\par\vskip18pt
    \normalsize\elsauthors\par\vskip10pt
    \footnotesize\itshape\elsaddress\par\vskip36pt
    \end{center}%
  \gdef\thefootnote{\arabic{footnote}}%
  }
\makeatother

\usepackage{amsmath}
\usepackage{amssymb}
\usepackage{amsfonts}
\usepackage{graphicx}
\usepackage{verbatim}
\usepackage{bm}
\usepackage{mathbbol}
\usepackage{color}
\setcitestyle{compress}

\usepackage{mathtools}

\usepackage{enumitem}

\usepackage{graphicx,epsfig,amsfonts,amssymb}
\usepackage{bm}
\usepackage{times}
\usepackage{lipsum}
\usepackage{verbatim}

\usepackage{hyperref}
\hypersetup{
    colorlinks,
    citecolor=blue,
    filecolor=blue,
    linkcolor=blue,
    urlcolor=blue
}

\newcommand{\beg}{\begin{equation}}
\newcommand{\en}{\end{equation}}

 \newcommand{\lam}{\lambda}

\newcommand{\eref}[1]{Eq.~(\ref{#1})}
\newcommand{\re}[1]{(\ref{#1})}

\newcommand{\esref}[1]{Eqs.~(\ref{#1})}
\renewcommand{\Re}{\mathrm{Re}}
\renewcommand{\Im}{\mathrm{Im}}

\newcommand{\eps}{\varepsilon}
\renewcommand{\phi}{\varphi}

\newcommand{\dn}{\downarrow}
\newcommand{\up}{\uparrow}

\newcommand{\pp}{{\bm p}}
\newcommand{\qq}{{\bm q}}
\newcommand{\kk}{{\bm k}}
\newcommand{\ii}{{\bm i}}
\newcommand{\jj}{{\bm j}}

\newcommand{\GG}{\mathcal{G}}
\newcommand{\FF}{\mathcal{F}}

 \newcommand{\be}{\begin{equation}}
\newcommand{\ee}{\end{equation}}
\def\bea {\begin{eqnarray}}
\def\eea {\end{eqnarray}}

\begin{document}

\title{Breakdown of  the Migdal-Eliashberg theory    and a  theory of  lattice-fermionic Superfluidity}

\author{Emil A. Yuzbashyan}
\affiliation{Department of Physics and Astronomy, Rutgers University, Piscataway, New Jersey 08854, USA}
 
\author{Boris L. Altshuler}
\affiliation{Physics Department, Columbia University, 538 West 120th Street, New York, New York 10027, USA}

\begin{abstract}

We show that the Migdal-Eliashberg theory loses validity at a finite value $\lambda_c$ of the  electron-phonon coupling $\lambda$ regardless of the underlying   model Hamiltonian. The value of $\lambda_c$ is approximately between 3.0 and 3.7. The new phase that emerges at $\lambda>\lambda_c$ breaks the lattice
translational symmetry. Depending on the filling fraction and crystal symmetry,  it is an insulator or a Fermi liquid. Its characteristic feature is a gap or a pronounced depression of the fermionic density of states near the Fermi level. We establish the breakdown from within the Migdal-Eliashberg theory by demonstrating that the normal state specific heat is negative for $\lambda\ge 3.7$ and  the quasiparticle lifetime vanishes in the strong coupling limit.
At fixed $\lambda>\lambda_c$, the transition to the new phase occurs at a critical temperature  higher than the superconducting transition temperature. In addition, there is a first order phase transition between the new phase and the superconducting state as we vary $\lambda$ across $\lambda_c$ at fixed temperature. We put forward a new theory --  lattice-fermionic theory of superfluidity -- that bridges the gap between the Migdal-Eliashberg approach and   the physics at stronger coupling. At small $\lambda$, our theory reduces to the Migdal-Eliashberg theory and,  past $\lambda_c$, it describes the new phase and a range of other phenomena.

\end{abstract}

\maketitle

\section{Introduction}

Migdal-Eliashberg theory~\cite{migdal,eli1st} is the principal theoretical framework for understanding properties of the normal and superconducting states
in metals   determined by boson  mediated electron-electron interactions. It is a time-dependent mean-field theory that makes accurate
quantitative predictions for a wide range of materials of the superconducting transition temperature,  quasiparticle gap, and most other thermodynamic and dynamical observables, many of which are beyond the Bardeen-Cooper-Schrieffer  (BCS) theory of superconductivity~\cite{bcs}. On a technical level, the Migdal-Eliashberg theory in its simplest formulation comes down to two coupled   self-consistency equations, known as the Eliashberg equations,
for the normal, $\Sigma(i\omega_n)$, and anomalous,  $\Phi(i\omega_n)$, self-energies that are functions of the fermionic Matsubara frequency $\omega_n=\pi T(2n+1)$.

The main open question in  the Migdal-Eliashberg theory is its status at strong \textit{renormalized} (actual) electron-phonon coupling $\lam$. This is the question  we address in the present paper. We prove that this theory  looses validity for $\lam>\lam_c$ irrespective of the underlying electron-phonon model, where
\beg
3.0\lesssim \lam_c\lesssim 3.7.
\label{range_lamc_intro}
\en
The smoking gun evidence of the breakdown is  negative  specific heat of the Migdal-Eliashberg normal state  indicating that  it is thermodynamically unstable~\cite{landau}.  Further evidence is the quasiparticle decay rate $\Gamma=\lam\pi T$ that is much larger than the temperature $T$ at strong coupling and diverges in the limit $\lam\to\infty$. To address the physics beyond $\lam_c$, we put forward a new theory -- lattice-fermionic theory of superfluidity in which the lattice and fermions are closer intertwined. At small $\lam$ it reduces to the Migdal-Eliashberg theory and past $\lam_c$ it describes the new phase. 

We  find that it is the interaction of electrons near the Fermi surface mediated by quantum fluctuations of the lattice that makes the specific heat negative within the Migdal-Eliashberg theory. The reason is that this theory misses an abrupt reconstruction of the electronic band structure and  formation of  new bound states   of the low energy electrons and quantum phonons that occur at $\lam=\lam_c$.  This   is assisted by static distortions of the lattice -- classical   phonons. Such  lattice distortions are generally thought to occur at strong electron-phonon interaction~\cite{millis,roland,alexandrov,meyer,capone,scalapino,esterlis}. We  show that  they are present already when $\lam$ is   finite  and the Fermi energy is the largest energy scale in the problem. We also find that vanishing of the quasiparticle lifetime in the limit $\lam\to\infty$ is entirely due to thermal fluctuations of  these distortions.  At the same time,  we will see that quantum rather than classical phonons are the main culprits in the breakdown of the Migdal-Eliashberg theory.

Consider for simplicity the Holstein model of electrons moving on a lattice and interacting with ions. The model assumes  ions are independent harmonic oscillators of mass $M$ and spring constant $K$ and takes the Coulomb interaction between electrons and ions to be
\beg
H_\mathrm{int}=\sum_{\bm i} (\alpha x_\ii) n_{\bm i},
\label{holsteinHint_intro}
\en
where $x_\ii$ is the displacement of the ion at site $\ii$  from its equilibrium position  and $n_\ii$ is the number of electrons at this site.
The dimensionless electron-phonon coupling constant is defined as
\beg
\lam=\frac{\nu_0\alpha^2}{K},
\label{lam_intro}
\en
where $\nu_0$ is the density of electronic states at the Fermi level per lattice site per spin projection. Note that we took $K$ to be the renormalized spring constant.

It is easiest to understand the physics at strong coupling  in the limit $\lam\to\infty$. Equation~\re{lam_intro} shows that this is   the free ion limit $K\to 0$. The oscillation frequency $\Omega=\sqrt{K/M}$ vanishes   together with $K$. Any finite temperature is much larger than $\Omega$. High temperature excites lattice oscillators  to states with large quantum numbers making them essentially classical (classical phonons). Integrals over momenta of classical oscillators in the partition function decouple from the remaining degrees of freedom. 
 We are left with the kinetic energy of the electrons plus \eref{holsteinHint_intro}, where $x_\ii$ are classical variables. The elastic energy $\sum_\ii K x_\ii^2/2$ vanishes in the strong coupling limit. The only role of the oscillators  now
 is to provide a statistically  distributed on-site potential $V_\ii=\alpha   x_\ii$ for the electrons. The spatial average of
$  x_\ii$ couples to the total electron number only and we absorb it into the chemical potential. 

In the mean-field approximation, the problem reduces to finding a nonuniform on-site potential $V_\ii$ for the electrons that minimizes their free energy. At $K=0$, the energy can be lowered indefinitely. At $K=\infty$ the only solution is $x_\ii =0$, since having nonzero $x_\ii$ costs infinite elastic energy. As we lower $K$, at a certain $K_c$,  $x_\ii$ become nonzero breaking  lattice translation symmetry  similarly to  the Peierls distortion~\cite{pouget,acta} and generating a frozen on-site potential $V_\ii=\alpha x_\ii$ for the fermions.  This
potential modifies the fermionic band structure.  Since we  take the Fermi energy to be much larger than any other characteristic energy,   the only possible relevant   modification   is a pronounced depression of the fermionic density of states near the Fermi level. We show that at least for certain system parameters a hard gap opens triggering a metal-insulator transition.

Migdal-Eliashberg theory assumes translational invariance. For example, electron and phonon Green's functions depend only on coordinate differences. This implies  thermal averages of   classical ion displacements  are zero, $\langle  x_\ii\rangle=0$. However,  fluctuations of $x_\ii$ are not.     $V_\ii= \alpha x_\ii$  is then equivalent to  disorder potential and thermal averaging to disorder averaging. Using the standard expression for  disorder averaged quasiparticle decay rate in a random potential, we find  $\Gamma=\lam \pi T$,
which coincides with the prediction of the Migdal-Eliashberg theory.  

We see that the divergence of the quasiparticle decay rate in the limit $\lam\to\infty$ is entirely due to classical phonons.
In contrast,  negative quasiparticle specific heat   cannot be explained in this way. Moreover, we will see that these   phonons cancel out from the Migdal-Eliashberg free energy  altogether. Therefore, even though classical phonons facilitate the abrupt change in the fermionic band structure, the breakdown of the theory at finite $\lam$  occurs only due to strong electron-electron interactions mediated by quantum fluctuations of the lattice, i.e., by quantum phonons.

\begin{figure}[htb]
\centerline{\includegraphics[width=0.9\columnwidth]{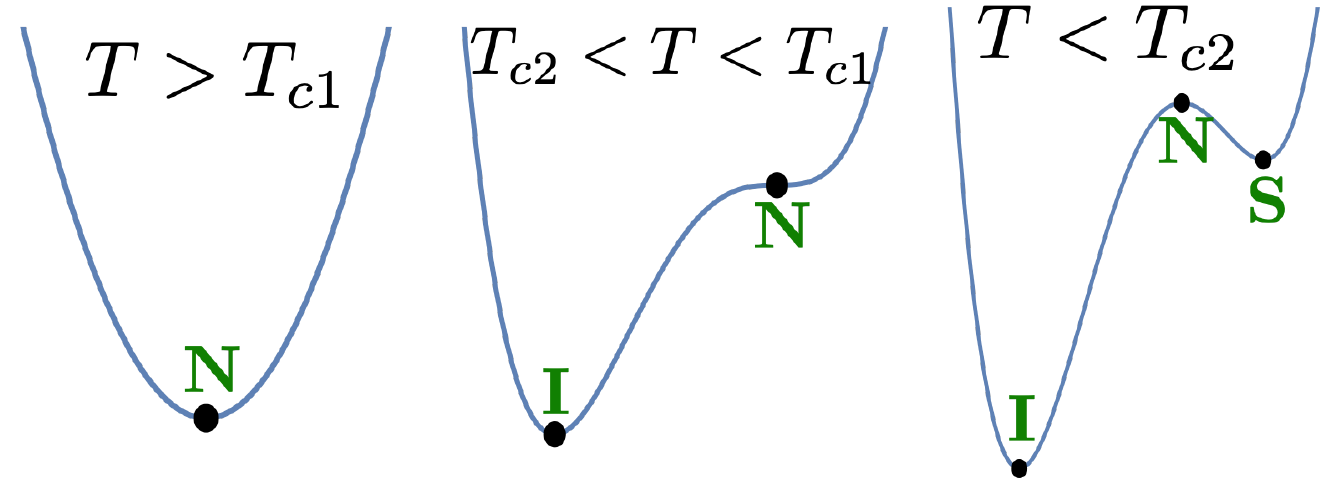}}
\caption{Free energy profile of the electron-phonon system after the Migdal-Eliashberg theory breaks down, $\lam>\lam_c$. New order emerges [insulator (I) in this case] below $T_{c1}$  out of the normal (N) state. Superconducting (S) stationary point appears at a lower temperature $T_{c2}$.  At temperatures just below $T_{c2}$, this stationary point must be higher in energy than the insulator by continuity. }
\label{shape_intro}
\end{figure}

Above arguments imply that a new order emerges in the electron-phonon system at strong coupling. Fix $\lam>\lam_c$. The transition to the new order
occurs at a critical temperature $T_{c1}$ above the superconducting transition temperature $T_{c2}$, since the heat capacity turns negative above $T_{c2}$.  We illustrate this in Fig.~\ref{shape_intro} where we show schematically the evolution of the free energy profile with $T$.  At very large $T$, we have a classical gas of electrons and phonons. Below $T_{c1}$ the new phase emerges, which we take to be an insulator for concreteness. At $T_{c2}$ the superconducting   stationary point appears, because the nontrivial solution of the Eliashberg equations exists at low enough $T$ for any
$\lam$ and corresponds to a stationary point of the free energy~\cite{spinchain}. Near its inception, the superconducting state is a local minimum or a saddle point, because the insulating minimum is already much below the normal state, see Fig.~\ref{shape_intro}. This means in particular that there is a first order transition between the superconductor and insulator as a function of $\lam$  at temperatures just below $T_{c2}$.

To describe the entire phase diagram of the electron-phonon system, we propose a new theory, which  we dubbed the theory of lattice-fermionic superfluidity.  The main idea is to incorporate the classical part of the phonon field into the single-particle Hamiltonian for fermions as an adjustable potential. This generally breaks the lattice translational symmetry. We treat
the boson-mediated interaction in particle-particle and particle-hole channels in saddle point approximation as in the Migdal-Eliashberg theory, except now the normal and anomalous self-energies $\Phi$ and $\Sigma$ depend on the single-particle state. The end result   is a set of four coupled equations that self-consistently determine classical  displacements of the oscillators from their equilibria, single-electron states and energies, and the fields $\Phi$ and $\Sigma$. This theory reproduces the Migdal-Eliashberg theory at $\lam<\lam_c$ and
the polaron formation theory~\cite{kabanov} in the adiabatic limit $M\to\infty$. It continues to work past $\lam_c$ and captures at least some of the new physics that emerges at strong coupling.

 In the above discussion, it is crucial  to distinguish   the  \textit{renormalized} electron-phonon coupling $\lam$
  and the  \textit{bare} coupling $\lam_0$.    Suppose $\Omega_0=\sqrt{K_0/M}$ is the bare frequency of Holstein phonons. Migdal and Eliashberg found that within standard electron-phonon models, such as the Fr\"olich or Holstein Hamiltonian, electrons strongly renormalize  the phonons, so that the renormalized phonon frequency is~\cite{migdal,eli1st,agd}
\beg
\Omega\approx \Omega_0\sqrt{1-2\lam_0}.
\label{Omega_renorm}
\en
 This formula predicts a lattice instability at $\lam_0\approx 0.5$ (lattice vibration frequencies become imaginary) restricting the domain
of applicability of the Migdal-Eliashberg theory to $0\le \lam_0\lesssim 0.5$, as Migdal and Eliashberg both note~\cite{migdal,eli1st,factor2}. Their conclusion that the theory does not work for $\lam_0\gtrsim 0.5$ has since been verified and elaborated upon by many other 
studies~\cite{millis,roland,alexandrov,meyer,capone,scalapino,esterlis}. However, it is important to emphasize that this does not necessarily violate
 Migdal's theorem~\cite{migdal,eli1st}, which says that quadratic fluctuations of the fermionic fields $\Sigma$ and $\Phi$ around the Eliashberg minimum of the free energy (point S in Fig.~\ref{shape_intro}) are small as long as the Fermi energy is sufficiently large~\cite{meaningmigdal}. We just have to keep in mind that this theorem applies only at the Eliashberg stationary point and not at other points, such as the insulating minimum I in  Fig.~\ref{shape_intro}. Because of the above lattice instability,  one of the main assumptions of the theory as formulated by Migdal and later Eliashberg is that $\lam_0$ is smaller than and not too close to 0.5, see also p. 182 of Ref.~\cite{agd}. This point is often overlooked in the literature and attempts are made to study the Migdal-Eliashberg theory outside of this interval of $\lam_0$. The finding that the theory does not work for such $\lam_0$ is not  news, but was  known already to Migdal and Eliashberg.

 The true question  therefore is not whether the   theory  stops working beyond $\lam_0\approx0.5$, but if there is an upper bound on the \textit{renormalized} electron-phonon coupling $\lam$. In terms of the electron-phonon interaction energy constant, $g^2=\nu_0\alpha^2 M^{-1},$ the renormalized coupling~\re{lam_intro} reads $\lam=g^2/\Omega^2$. Similarly, the bare electron-phonon coupling is
 $\lam_0=g^2/\Omega_0^2$.   Equation~\re{Omega_renorm} then implies
\beg
\lam=\frac{\lam_0}{1-2\lam_0}.
\label{lambda_renorm}
\en
We see that $\lam$ varies from 0 to $+\infty$ within the domain of applicability, $0\le \lam_0\lesssim 0.5$, of the Migdal-Eliashberg theory. This seems to suggest
that arbitrarily large values of $\lam$ are attainable~\cite{andrey_validity}. Since at strong coupling $T_c/\Omega\approx 0.183\sqrt{\lam}$~\cite{allendynes}, this would imply  unbounded $T_c$ in units of  the characteristic phonon frequency.   

 About a decade after Migdal's work Brovman and Kagan  realized that the above lattice instability  is in fact merely an artifact of conventional electron-phonon models~\cite{kagan,geilikman}.   Such models postulate certain lattice vibration spectra, e.g., acoustic or optical phonons, and a certain form of electron-phonon interaction. These phonon spectra are already a product of electron-lattice interactions and their further renormalization by these interactions is unwarranted.  In the proper (adiabatic) perturbation theory in the ratio of the electron to ion mass, we  start by solving for the energy of the electrons for given  ion displacements. We then combine this energy with the Coulomb interaction between the ions to solve the lattice vibrational problem and determine  phonon frequencies. 
 
The zeroth order Hamiltonian for ions has ions interacting via unscreened Coulomb interactions -- ionic plasma, where ions oscillate with the plasma frequency.  Electrons renormalize these plasma oscillations converting them, for example, into acoustic phonons with no lattice instability along the way.   Modern state of the art simulations observe this    in ``an approximation free way"~\cite{tupitsyn}.  Conventional models on the other hand start with an ansatz for the phonon dispersion and electron-phonon interaction. Consider, for example, a model  of electrons interacting with acoustic phonons. It is this interaction
that renormalizes the phonon spectrum within this model leading to the above lattice instability. However, such a secondary renormalization is double counting as we already renormalized lattice vibrations once to obtain acoustic phonons. Because of this the consensus in the community has been that one should not   renormalize the phonons within the Migdal-Eliashberg theory, but  instead supplement the theory by experimentally measured phonon frequencies~\cite{mitrovic,2008review}. 

We adopt the same approach in this paper. We keep the phonon spectrum arbitrary and show that the Migdal-Eliashberg theory loses
validity at a certain finite $\lam_c$ independently of the phonon dispersion law and the momentum dependence of the electron-phonon matrix element, i.e., independently of the underlying electron-phonon Hamiltonian. This is possible because the strong coupling limit of this theory is universal. Similar to its weak coupling limit (BCS theory), there is a single energy scale in this limit~\cite{combescot}. Prior studies  mix up the above lattice instability, which is outside of the domain of  applicability of  the Migdal-Eliashberg theory, with its true breakdown within its domain. Many of them are model-dependent and do not make the necessary distinction between the bare and renormalized electron-phonon coupling constants. Most importantly, they do not  eliminate  the possibility that the theory remains valid for all $\lam$, including $\lam=\infty$. It is also important to note that the  mechanism of the breakdown we discussed above, while also accompanied by a   lattice  distortion, is   unrelated to the lattice instability due to the artificial phonon softening at $\lam_0\approx 0.5$. Indeed,  no such softening takes place in our mechanism.  

The paper is organized as follows: In Secs.~\ref{eli_sec} and \ref{spinsect}, we review our previous work~\cite{spinchain} where we
derived the  quasiparticle  free energy within the Migdal-Eliashberg theory and mapped it to a classical spin chain. We also introduce
models we  employ in this paper and discuss alternative forms of the Eliashberg equations and the strong coupling limit of the theory. In Sec.~\ref{heatsect}, we establish that  Migdal-Eliashberg theory loses validity for $\lam>\lam_c$, where $\lam_c\lesssim 3.7$. The normal state heat capacity becomes negative for $\lam>3.7$ and  quasiparticle decay rate is much larger than the temperature. In contrast, the superconducting state is free from such pathologies. We further show in Sec.~\ref{heatsect} that at strong coupling the low energy part of the quasiparticle spectrum of the superconductor consists of narrow bands of width of the order of the phonon frequency $\Omega$. At high energies, the spectrum is continuous with no gaps. The specific heat is  positive at all $T$ in the supeconducting state and the quasiparticle decay rate is negligible. We develop a
simple qualitative picture of the breakdown that explains the above pathologies of the normal state in Sec.~\ref{qualitative_sect}. 
 In Sec.~\ref{new_order_sect}, we discuss new order that emerges at $\lam_c$ and its implications for the electron-phonon system.
 In Sec.~\ref{compare_sect}, we compare our and previous studies  of the Migdal-Eliashberg theory.
 Sec.~\ref{static_sect} addresses the role of  classical phonons in the breakdown and in Sec.~\ref{adiabatic} we consider the adiabatic limit, $M\to\infty$, that reveals their role   especially clearly. In Sec.~\ref{new_theory_sect}, we present our theory of lattice-fermionic superfluidity that remains valid after the Migdal-Eliashberg theory breaks down and  accommodates  new phases  emerging at stronger coupling. In concluding section, we summarize and discuss open questions and some of the implications of our study, such as an upper bound on the superconducting $T_c$.

\section{Free energy  and Eliashberg equations}
\label{eli_sec}

We begin with the description of two electron-phonon models that we use -- the Holstein model and a  more general model
with arbitrary phonon spectrum and momentum dependent electron-phonon interaction. We then review the results of our earlier work  where we derived the free energy density functional for the fermionic subsystem for these models. The stationary point equations of this free energy are standard Eliashberg equations, which we also review.

\subsection{Models}

We employ two models in this paper. The first one is the Holstein model (dispersionless phonons) with  an \textit{arbitrary} hopping matrix and an onsite potential,
\beg
H=\sum_{\bm i \bm j \sigma} h_{\bm i\bm j} c^\dagger_{\bm i\sigma} c_{\bm j \sigma}+  \sum_{\bm i} \left[\frac{p_\ii^2}{2M} + \frac{K_0 x_\ii^2}{2}\right] +
\alpha \sum_{\bm i} n_{\bm i} x_\ii,
\label{holsteinH}
\en 
where $\ii$ and $\jj$ label  the lattice sites, $h_{\ii\jj}$ are the matrix elements of an  arbitrary single-electron Hamiltonian $\hat h$, $c^\dagger_{\bm i \sigma}$ and  $c_{\bm i \sigma}$  are creation and annihilation operators for an electron on site $\ii$ with spin projection $\sigma$,  $n_\ii =\sum_\sigma c^\dagger_{\ii\sigma} c_{\ii\sigma}$ is the fermion occupation operator, and $p_\ii$ and $x_\ii$ are ion momentum and position operators. The \textit{bare} phonon frequency is $\Omega_0=\sqrt{K_0/M}$.  

The second model is a more general   Hamiltonian describing electrons interacting with dispersing phonons,
\beg
\begin{split}
H=&\sum_{\pp \sigma} \xi_\pp c^\dagger_{\pp\sigma} c_{\pp\sigma} +  \sum_\qq {\omega_{0}(\qq) } b^\dagger_\qq b_\qq\\
&+\frac{1}{\sqrt{N}} \sum_{\pp \qq\sigma} \frac{\alpha_{\qq}}{\sqrt{2M\omega_0(\qq)}} c^\dagger_{\pp+\qq \sigma} c_{\pp\sigma} \left[ b^\dagger_{-\qq} + b_\qq\right],
\end{split}
\label{frol1}
\en
where $M$ is the ion mass and $N$ is the number of lattice sites. The phonon spectrum $\omega_{0}(\qq)$ and the electron-phonon interaction $\alpha_{\qq}$ are largely arbitrary, except that we will assume for simplicity that both depend on the magnitude of the momentum only, $\omega_{0}(\qq)=\omega_{0}(q)$ and  $\alpha_{\qq}=\alpha_q$. Both $h_{\ii\jj}$ and $\xi_\pp$ contain the chemical
potential $\mu$ as we include the $\mu N_\mathrm{f}$ term into the Hamiltonians, where $N_\mathrm{f}$ is the total fermion number. 
 
 \subsection{Free energy functional}
 \label{free_en_sub}
 
 In the first paper~\cite{spinchain} in our series of four papers~\cite{spinchain,meaningmigdal,retardation} on the Migdal-Eliashberg theory, we derived the free energy   functional (effective action) for spatially homogeneous states of the system for both
 above Hamiltonians. The idea is to integrate out  phonons in the path integral and then  decouple   resulting effective fermion-fermion interaction with three Hubbard-Stratonovich fields
$\Phi$, $\Sigma_\up$ and $\Sigma_\dn$. Next, we integrate out the fermions, obtain an effective action  in terms of the \textit{Eliashberg fields} $\Phi$, $\Sigma_\up$ and $\Sigma_\dn$, and look for  stationary points where these   fields are spatially uniform and depend on the time difference only.  We work in the regime where the Fermi energy is the largest  energy in the problem, much larger than characteristic interaction and phonon energies. This implies that the single-fermion spectrum is particle-hole symmetric at relevant energies and  we also assume time reversal symmetry. 

The above steps  and setup lead to
 the following expression for the free energy  of the system per site~\cite{grand}:
\beg
\begin{split}
f= \nu_0 T^2\sum_{nl}\left[ \Phi_{n+l}^* \Lambda_l \Phi_{n} + \Sigma_{n+l} \Lambda_l  \Sigma_{n}    \right]\\
-2\pi \nu_0 T\sum_{n} \sqrt{ (\omega_n+\Sigma_{n} )^2 +| \Phi_{n} |^2 }.
\end{split}
\label{feH1}
\en
Here $\nu_0$ is the density of states at the Fermi energy per site per spin projection. The  field $\Phi_n\equiv\Phi(i\omega_n)$ is complex and $\Sigma_n\equiv\Sigma(i\omega_n)$ is real. Both fields are functions of the fermionic Matsubara frequency $\omega_n=\pi T(2n+1)$. Particle-hole symmetry implies that $|\Phi_n|$ is even and $\Sigma_n$ is odd in $\omega_n$.  At the stationary point, these fields equal the anomalous and normal self-energies. The effective action is $S_\mathrm{eff}= N  f/T$ with $N$ being the number of lattice sites. At its minimum \eref{feH1} gives the grand potential of the system in the thermodynamic limit, but we colloquially   refer to it as the free energy. More generally, $e^{-Nf/T}$ determines the weight of a given field configuration in the partition sum.

  The quantity $\Lambda_l$ in \eref{feH1} is the Fourier transform of $1/\lambda(\tau)$ at bosonic Matsubara frequency $\omega_l=2\pi T l$, where $\lam(\tau)$ is the effective   electron-electron interaction in the imaginary time domain. It is more practical to specify 
 $\lam(\omega_l)$ -- the Fourier transform of $\lam(\tau)$ to the Matsubara frequency domain. 
  For the Holstein model, we have
\beg
\lambda(\omega_l)= \frac{  g^2}{\omega_l^2+\Omega^2},\quad g^2=\nu_0 \alpha^2 M^{-1}.
\label{Holstein_lambda}
\en
 For phonons with dispersion 
\beg
 \lambda(\omega_l)=\frac{1}{2p_F^2} \int_0^{2p_F}\!\!\!   \frac{ g_q^2 q dq}{\omega_l^2+\omega_{q}^2},\quad g_q^2=\nu_0  |\alpha_{q}|^2  M^{-1},
 \label{int_disp}
 \en
 where $p_F$ is the Fermi momentum. Here $\Omega$ and $\omega_{q}$ are the \textit{renormalized} phonon frequencies, not to be confused with  bare frequencies $\Omega_0$ and $\omega_0(q)$. The last expression for $\lam(\omega_l)$ is for a spherical Fermi surface in $d=3$ dimensions, but it is straightforward to extend it to any   $d\ge 2$. 
 
 As usual, we  define the dimensionless electron-phonon coupling constant as $\lambda=\lambda(\omega_l=0)$. Then,
 \beg
 \begin{split}
 \lambda=&\frac{g^2}{\Omega^2}=\frac{\nu_0\alpha^2}{K}, \quad\mbox{Holstein model,}\\
 \lambda= &\frac{1}{2p_F^2}  \int_0^{2p_F}  \frac{  g_q^2 q dq}{\omega^2_{q}},  \quad \mbox{dispersing phonons,}\\
 \end{split}
 \label{lam}
 \en
 where $K$ is the renormalized spring constant. It is also convenient to introduce $\alpha$, $g$, $K$ and $\Omega$ for dispersing phonons
 as the following averages:
 \beg
 \begin{split}
 \alpha^2&\equiv \frac{1}{2p_F^2}  \int_0^{2p_F} |\alpha_{q}|^2 q dq,\quad g^2\equiv\nu_0 \alpha^2 M^{-1},\\
 \Omega^2&\equiv\frac{g^2}{\lam},\quad K\equiv\frac{\nu_0\alpha^2}{\lam}.
 \end{split}
 \label{averages}
 \en
 Often we will consider the strong coupling limit $\lam\to\infty$, which is equivalent to $\Omega\to 0$ or $K\to0$.

 \subsection{Eliashberg equations}
   
 The stationary point equations for the free energy~\re{feH1}, are the well-known \textit{Eliashberg equations}~\cite{eli1st},
\begin{subequations}
\begin{eqnarray}
\Phi_n= \pi  T \sum_m  \lambda_{nm}\frac{ \Phi_m}{ \sqrt{(\omega_m+\Sigma_{m} )^2 +| \Phi_{m} |^2} },\label{Phieq} \\ 
\Sigma_n= \pi  T \sum_m  \lambda_{nm}  \frac{\omega_m+ \Sigma_m}{ \sqrt{(\omega_m+\Sigma_{m} )^2 +| \Phi_{m} |^2}}, \label{Sigmaeq}
\end{eqnarray}
\label{elifamiliar3}
\end{subequations}
where 
\beg
\lam_{nm}=\lambda(\omega_n-\omega_m),\quad \lambda_{nn}=\lambda(0)=\lambda.
\label{lambda_nm}
\en
Note that $\lam_{nn}$ diverges in the strong coupling limit.

It is convenient to introduce new variables -- complex $F(\tau)$ and real $G(\tau)$ defined as
\beg
\Phi(\tau)=\pi \lambda(\tau) F(\tau),\quad \Sigma(\tau)=\pi \lambda(\tau) G(\tau).
\en
In frequency representation we have
\beg
\begin{split}
\Phi_n=\pi T\sum_m  \lambda_{nm} F_m,\quad
 \Sigma_n=\pi T\sum_m  \lambda_{nm} G_m,
 \end{split}
 \label{varchangeR}
\en
and \eref{elifamiliar3} becomes
\beg
\begin{split}
 F_n=\frac{ \Phi_n}{ \sqrt{(\omega_n+\Sigma_{n} )^2 +| \Phi_{n} |^2} },\\
G_n = \frac{\omega_n+ \Sigma_n}{ \sqrt{(\omega_n+\Sigma_{n} )^2 +| \Phi_{n} |^2}}.
\end{split}
\label{fg}
\en
On the stationary point, the fields $F_n$ and $G_n$ correspond to the anomalous and normal Green's functions integrated over the single-particle energy~\cite{spinchain}.   Parity properties of $\Sigma_n$ and $|\Phi_n|$ imply that $G_n$ is  odd and $|F_n|$ is even.

Importantly, it is possible to rewrite the Eliashberg equations~\re{elifamiliar3} so as to eliminate the $m=n$ terms that diverge in the strong coupling limit~\cite{andrey_validity,spinchain}. The new equations have the same form
\begin{subequations}
\begin{eqnarray}
\Phi'_n= \pi  T \sum_{m\ne n}  \lambda_{nm}\frac{ \Phi'_m}{ \sqrt{(\omega_m+\Sigma'_{m} )^2 +| \Phi'_{m} |^2} },\label{Phieq1} \\ 
\Sigma'_n= \pi  T \sum_{m\ne n}  \lambda_{nm}  \frac{\omega_m+ \Sigma'_m}{ \sqrt{(\omega_m+\Sigma'_{m} )^2 +| \Phi'_{m} |^2}}, \label{Sigmaeq1}
\end{eqnarray}
\label{elifamiliar311}
\end{subequations}%
where
\beg
\begin{split}
\Phi'_n=\pi T\sum_{m\ne n}  \lambda_{nm} F_m,\quad
 \Sigma'_n=\pi T\sum_{m\ne n}  \lambda_{nm} G_m.
 \end{split}
 \label{varchange11}
\en
Troublesome  $m=n$ terms are now absent both from  the new equations and from the ``reduced" self-energies $\Phi'_n$ and $\Sigma'_n$.

Eliashberg equations generally have more than one solution, e.g., at $T=0$ there is a solution with $\Phi_n=0$ and a solution
with $\Phi_n\ne0$. Moreover, we established in Ref.~\cite{spinchain} that new ``spin flip" solutions   emerge at $\lam\gtrsim 1$.
However, most important for us here is the solution with the lowest free energy, which we dub the \textit{Eliashberg stationary point}. This stationary point is the global minimum when the Migdal-Eliashberg theory is a valid description of the system and is a saddle point or a local   minimum otherwise.

 \section{Mapping to a spin chain}
 \label{spinsect}

 This section concludes the summary of our previous results that  we will use  to demonstrate the breakdown of the Migdal-Eliashberg theory at strong coupling from within the theory itself. The main result reviewed here is that the free energy functional maps to a classical Heisenberg spin chain. 
 Sites of the chain are fermionic Matsubara frequencies $\omega_n$ and the components of classical spin $\bm S_n$   are energy-integrated normal and anomalous Green's functions, $S_n^z=G_n$ and $S_n^+= F_n$, where $S_n^+\equiv S_n^x + i S_n^y$.
 
 Indeed, observe that \eref{fg} implies a constraint on the variables  $G_n$ and $F_n$,
\beg
G_n^2+|F_n|^2=1.
\label{constraint}
\en
Therefore we can treat these variables as three components of a \textit{classical spin} $\bm S_n$ of unit length, $\bm S^2_n=1$,
\beg
S_n^z=G_n,\quad S_n^x=\Re (F_n),\quad S_n^y=\Im (F_n).
\label{sgf}
\en
It follows from \eref{fg} that,
\beg
F_n\Phi_n^*+G_n(\omega_n +\Sigma_n)= \sqrt{(\omega_n+\Sigma_{n} )^2 +| \Phi_{n} |^2}.
\label{sqrt}
\en
This allows us to rewrite the free energy density given by \eref{feH1} as
\begin{align}
f=&\nu_0 T H_s,  \mbox{ where,}\label{Hsdef}\\
H_s =&-2\pi \sum_n \omega_n S_n^z-\pi^2 T\sum_{nm} \lambda_{nm}({\bm S_n}\cdot {\bm S_m}-1).
\label{spinh}
\end{align}
We interpret  $H_s$  as a Hamiltonian  of an open classical Heisenberg spin chain in an inhomogeneous ``Zeeman magnetic field". The positions of the spins are fermionic Matsubara frequencies $\omega_n$. Spin-spin interactions are ferromagnetic and fall off at large ``distance" as $\lambda_{nm}\propto (\omega_n-\omega_m)^{-2}$. The magnetic field is linear in the position of the spin and goes to $\pm\infty$ as $\omega_n\to\pm\infty$.  Eliashberg equations~\re{elifamiliar3} are spin equilibria conditions that enforce parallel alignment of each spin and the effective magnetic field acting on it (Zeeman field plus the field from other spins).

In particular,  for the Holstein model substituting \eref{Holstein_lambda} into \eref{spinh} we obtain
\beg
H_s=-2\pi \sum_n \omega_n S_n^z-\pi^2 Tg^2\sum_{nm} \frac{ {\bm S_n}\cdot {\bm S_m} -1 }{(\omega_n-\omega_m)^2+\Omega^2}.
\label{schol}
\en
In general,  the spin chain representations~\re{spinh} and \re{schol} of the free energy are guaranteed to work only at its stationary points, because we used \esref{constraint} and \re{sqrt} that derive from the stationary point equations to obtain them. It is also important
to keep in mind that $f$ is the contribution of the fermionic degrees of freedom (quasiparticles) to the  free energy. The total free energy is $f$ plus the free energy (grand potential) of noninteracting phonons.
See  Ref.~\cite{spinchain} for a comprehensive discussion of properties and consequences of the spin chain representation of the free energy.

\begin{figure}
	\centering
	 \includegraphics[width=0.85\columnwidth]{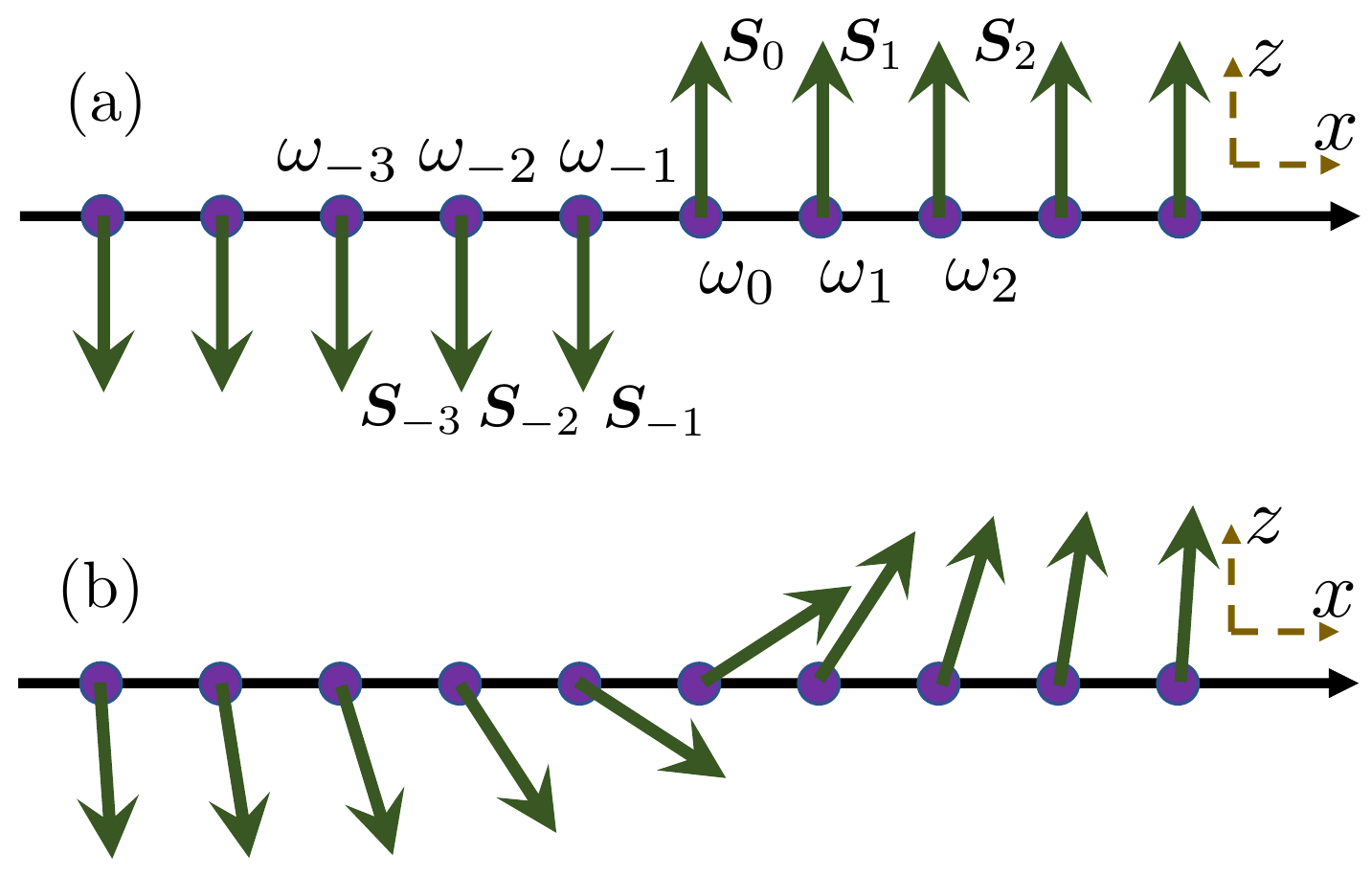}
	\caption{ Classical spin representation of the transition from (a) the normal to (b) superconducting state. As discussed  in the text,  the Migdal-Eliashberg theory maps to a classical Heisenberg spin chain. The positions of the spins are fermionic Matsubara frequencies $\omega_n$. Spin-spin interactions are purely ferromagnetic and the spins are subject to a Zeeman magnetic field $2\pi\omega_n$  along the $z$-axis.    In the superconducting state, spins acquire  $x$-components, which implies nonzero anomalous  Green's function. The sharp domain wall in the normal state is smeared in the superconducting state.  }
	\label{spinsfig}
\end{figure}

Now let us develop a mini-dictionary between the original language of superconductivity and the spin terminology.
 First of all, similar to the Anderson pseudospin description of the BCS theory of superconductivity~\cite{pseudo}, the superconducting transition translates into   softening of the domain wall as shown in Fig.~\ref{spinsfig}.   
This is a result of the competition between the Zeeman magnetic field and ferromagnetic interaction   in $H_s$. The spin configuration minimizing the  Zeeman term is $\bm S_n=\mbox{sgn} (\omega_n)  \hat {\bm z}$,  and the Zeeman field  necessarily  prevails  at large $|\omega_n|$, so that $\bm S_n\to \pm  \hat {\bm z}$ for $\omega_n\to\pm\infty$.  

Above the superconducting $T_c$, the anomalous averages vanish, $F_n=0$. According to the definition~\re{sgf} of the classical spin, this implies that all spins are parallel to the $z$-axis. From the behavior of $\bm S_n$ at large $\omega_n$ and by symmetry, it is then clear that the minimum energy spin texture is
\beg
\bm S_n=\mbox{sgn} (\omega_n)  \hat {\bm z}.
\label{normal_state}
\en
 This is the \textit{normal state} in the spin language. Its characteristic feature   is a sharp domain wall between $\omega_{-1}$ and $\omega_0$ with an abrupt   jump of   the $z$-component of spin from $S_{-1}^z=-1$ to $S_0^z=+1$, see Fig.~\ref{spinsfig}. 

Below $T_c,$ the anomalous averages are nonzero, i.e.,  the spins acquire $x$-components ($F_n$ can be made real in the spin chain ground state). In other words, the domain wall softens in superconducting states. Now the change in $S_n^z$  from $-1$ at $\omega_n=-\infty$ to $+1$ at $\omega_n=+\infty$ occurs gradually and the jump $S_{0}^z-S_{-1}^z <2$.  

 \subsection{Strong coupling limit}
 \label{strong_subsec}
 
 We will see in the next section that the Migdal-Eliashberg theory stops being a valid description of any physical system for    values of the renormalized electron-phonon coupling $\lam\ge \lam_c$, where $\lam_c\lesssim\lam_*\approx 3.69$. One may ask then, what is  the point in considering its strong coupling, $\lam\to\infty$, limit where the theory is unphysical.  The main point is that the answers for any observable  obtained with different underlying electron-phonon models, i.e., for different phonon spectra and electron-phonon matrix elements, converge in this limit -- the strong coupling limit of the Migdal-Eliashberg theory  is universal, see, e.g., Ref.~\cite{combescot}.
 
 For example, suppose we evaluated the specific heat for the Holstein model, $c_H(T,\lam)$, as a function of temperature and $\lam$. At $\lam=\infty$ the specific heat, $c_g(T,\lam)$, for a general electron-phonon model~\re{frol1} coincides with $c_H(T,\lam)$,  $c_g(T,\infty)=c_H(T,\infty)$. At large but finite $\lam$, $c_g(T,\lam)$ is close to $c_H(T,\lam)$ and we can make them arbitrarily close by increasing $\lam$. In particular, we will see that the value $\lam_*\approx 3.69$ obtains from the condition $\min[ c_H(T,\lam_*)]=0$ for $T>T_c$. The universality of the strong coupling limit implies that there is finite $\lam_c$ for the   general electron-phonon model as well. Moreover, since $\lam_*\approx 3.69$ is already quite large, the values of $\lam_*$ obtained for different models should be close to 3.69.

 The strong coupling limit   is equivalent~\cite{combescot,spinchain} to sending all (renormalized) phonon frequencies to zero, $\omega_q\to0$ and $\Omega\to 0$. Then the effective electron-electron interaction~\re{int_disp} becomes
 \beg
 \lam(\omega_l)=\frac{g^2}{\omega_l^2},
 \en
 where for the Holstein model $g^2=\nu_0 \alpha^2 M^{-1}$ as before, while for the dispersing phonon model~\re{frol1} the constant  $g^2$ is given by \eref{averages}.
   The free energy functional~\re{feH1} becomes in this limit~\cite{spinchain}
 \beg
H_s=-2\pi \sum_n \omega_n S_n^z-\pi^2 Tg^2\sum_{nm} \frac{ {\bm S_n}\cdot {\bm S_m}-1}{(\omega_n-\omega_m)^2}.
\label{schol1}
\en
We see explicitly that $H_s$ is independent of the underlying microscopic model except through a single constant $g$. Moreover, the $\lam\to\infty$ limit has another convenient property -- in this case the  expression~\re{schol1} for the free energy   holds at all points $(G_n, F_n)$  of the configuration space, unlike finite $\lam$, for which the spin chain representation~\re{spinh} applies only at  the stationary points of \eref{feH1}, see Ref.~\cite{spinchain} for more detail.

\section{Breakdown of the Migdal-Eliashberg theory}
\label{heatsect}

We present indisputable  evidence  of the breakdown of the Migdal-Eliashberg theory at strong coupling. The critical value of the   renormalized electron-phonon coupling  $\lam_c$ above which the theory  becomes unphysical lies in the interval $3.0\lesssim\lam_c\lesssim3.7$. We provide two pieces of such evidence. First, the normal state specific heat evaluated within this theory becomes negative above $T_c$
for $\lam\ge 3.7$ indicating that this state is thermodynamically unstable~\cite{landau}. 

Second,  quasiparticle lifetime  vanishes in the normal state as $\tau=[\mbox{Im} \Sigma(\omega)]^{-1}\approx(\pi\lam T)^{-1}\to 0$  when $\lam\to\infty$ signaling a complete breakdown of the quasiparticle picture.
 It indicates that the Migdal-Eliashberg theory no longer employs the correct zeroth order (in electron-phonon coupling) Hamiltonian for fermions. The true normal state   cannot be a  Fermi liquid with the  translational symmetry of the lattice anymore. We will see that this  short lifetime is entirely due to the thermal fluctuations of  static displacements of the ions from their equilibrium positions, see also Ref.~\cite{andrey_validity}.
 
 On the other hand, the behavior of these quantities in the superconducting state at strong coupling is diametrically opposite. The specific heat is positive for all $T\le T_c$ and   exhibits activated behavior  at low temperatures. Quasiparticle lifetime is  exponentially large   at $\lam=\infty$  and proportional to $\sqrt{\lam}$ at large but finite $\lam$. However, this does not mean  that the superconducting state predicted by this theory is ``out of the woods'' and indeed we will see in the next section that, at least for a range of temperatures below $T_c$, it is not the true thermal equilibrium  of the electron-phonon system. 
 
By construction  solutions of  the Eliashberg equations~\re{elifamiliar3}   are stationary points of the free energy functional for any coupling $\lam$. The above findings demonstrate that for $\lam > \lam_c$    none of these   stationary points is the true global minimum of the free energy, at least in a certain temperature range that includes temperatures both above and below $T_c$. In subsequent sections, we will see that this happens due to a phase transition that breaks the translational invariance of the lattice. This transition is    independent of  the lattice instability discussed in Introduction and does not rely on conventional electron-phonon models for its existence.  

\subsection{Normal state specific heat}
\label{normal_heat_sec}

We start by rederiving  the specific heat for the Holstein model within the Migdal-Eliashberg theory~\cite{grimvall} with the help of the spin chain Hamiltonian
\beg
H_s=-2\pi \sum_n \omega_n S_n^z-\pi^2 Tg^2\sum_{nm} \frac{ {\bm S_n}\cdot {\bm S_m}-1}{(\omega_n-\omega_m)^2+\Omega^2}.
\label{schol3}
\en
 By definition~\re{Hsdef} of the spin Hamiltonian, the free energy is $f=\nu_0 T H_s$. We saw in Sec.~\ref{spinsect} that in the normal state
\beg
S_n^z=\mbox{sgn} (\omega_n),\quad S_n^x=S_n^y=0.
\label{nsspins}
\en
Therefore the normal state free energy is
\beg
\begin{split}
f_\mathrm{n}=&-2\pi\nu_0 T \sum_n \omega_n \mbox{sgn} (\omega_n)\\
&-\pi^2\nu_0 T^2 g^2\sum_{nm} \frac{ \mbox{sgn} (\omega_n\omega_m)-1}{(\omega_n-\omega_m)^2+\Omega^2}.
\end{split}
\label{nsf}
\en
The first term on the right   is the temperature-dependent part of the free energy of noninteracting  electrons~\cite{zeta2},
 \beg
 f_0=-\frac{\pi^2 \nu_0 T^2}{3}.
 \label{nif}
 \en
Note that $\mbox{sgn} (\omega_n\omega_m)-1$ vanishes when $\omega_n$ and $\omega_m$ have the same sign and is equal to $-2$ otherwise. This observation allows us to rewrite the second term (interaction part of the free energy) as
\beg
f_\mathrm{int}=\nu_0 g^2\sum_{l=1}^\infty \frac{l}{l^2+a^2},\quad a\equiv\frac{\Omega}{2\pi T},
\label{fint}
\en
We also reduced the summation over $n$ and $m$ to a single sum over $l=n+m+1$ using $\omega_n-(-\omega_m)\propto (n+m+1)$ and taking into account that there are $l$ ways to choose $n$ and $m$ for a given value of $l$.

The sum in \eref{fint} is logarithmically divergent~\cite{nozeta}.   Nonetheless, let us write the summand as a sum of two simple fractions and use the following property of the digamma function $\psi(x)$:
\beg
\sum_{l=1}^\infty\left(\frac{1}{x+l}-\frac{1}{l}\right)=-\frac{1}{x}-\psi(x)-\gamma,
\en
where $\gamma$ is Euler's constant. We find,
\beg
\sum_{l=1}^\infty \frac{l}{l^2+a^2}= \mbox{Re}[\psi(ia)] +\sum_{l=1}^{l_0}\frac{1}{l} -\gamma,
\label{digamma}
\en
where we truncated the sum on the right hand side at $l_0$. For large $l_0$ the last two terms sum to $\ln l_0$ with an error of order $1/l_0$. Since $\Lambda=2\pi T l_0$ is a Matsubara frequency,  it is safe to replace these two terms with $\ln \frac{\Lambda}{2\pi T}$, where $\Lambda$ is the frequency cutoff. 

\begin{figure}[ht]
\begin{center}
\setlength{\unitlength}{10cm}
\begin{picture}(0.83, 0.628)(0,0)
   \put(0,0){\resizebox{0.82\unitlength}{!}{\includegraphics{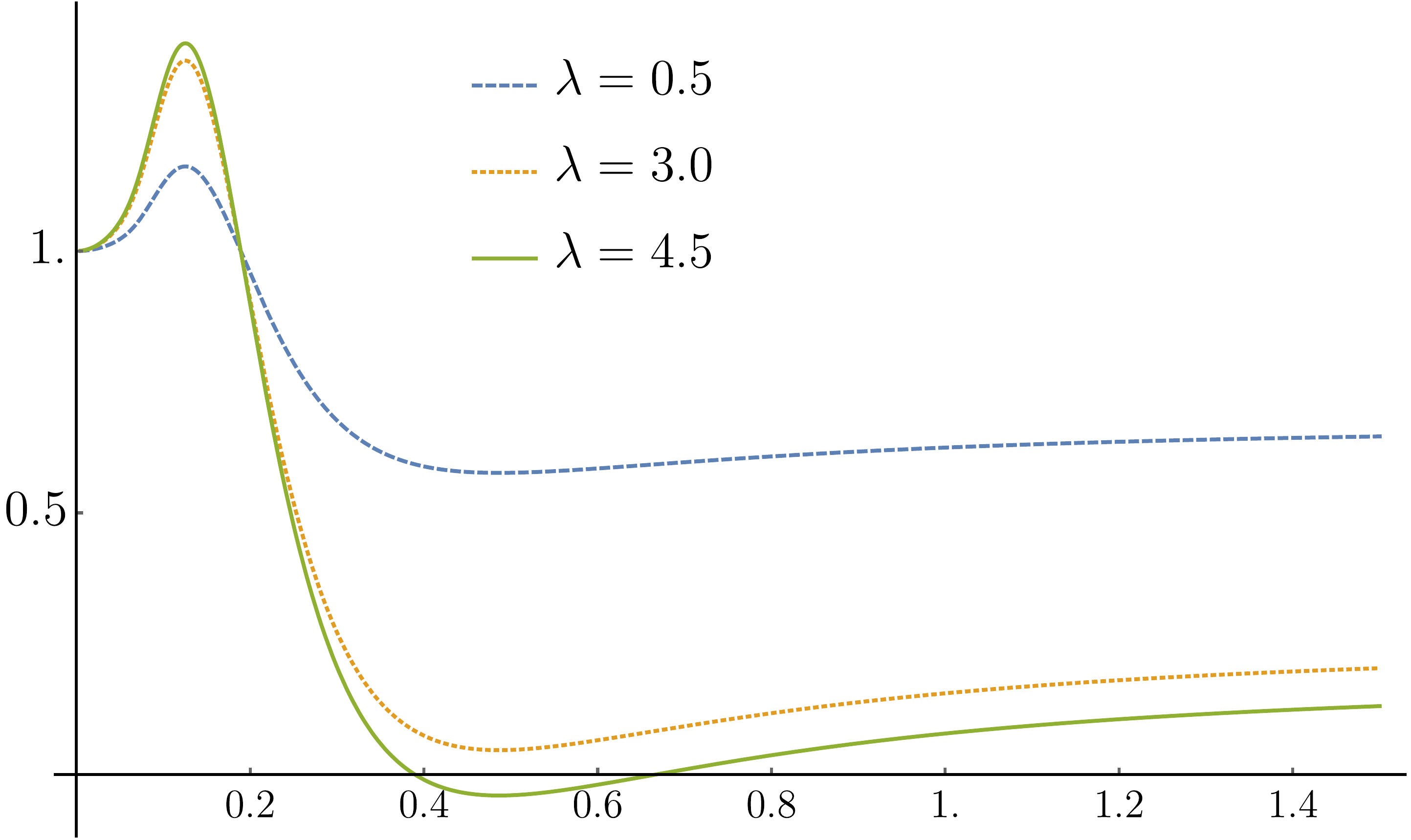}}}
    \put(0.84, 0.01){\makebox(0,0)[b]{ $\displaystyle \frac{T}{\Omega}$}}
   \put(-0.006,0.51){ {\makebox(0,0)[t]{  $ \displaystyle\frac{C_\mathrm{n}}{C_\mathrm{FL}}$}}}
\end{picture}
\end{center}
\caption{Normal state quasiparticle specific heat $C_\mathrm{n}$ in the Migdal-Eliashberg theory  as a function of the ratio $T/\Omega$ of  the temperature  to  Einstein phonon frequency   for three different values of   electron-phonon coupling  $\lam$. We normalize $C_\mathrm{n}$  by its low temperature  Fermi liquid asymptote $C_\mathrm{FL}=\gamma_0 T(1+\lam)$. Note that for $\lam=4.5$ the  specific heat is \textit{negative}  above the superconducting $T_c\approx 0.4\Omega$ signaling the \textit{breakdown} of the Migdal-Eliashberg theory at strong coupling. Since the $\lam\to\infty$ limit of this theory is universal, it breaks down irrespective of the   underlying electron-phonon model. }
\label{c}
\end{figure}

Note that $\Lambda$ affects only the temperature-independent part of the free energy that we are not attempting to evaluate anyway. This part is the ground state energy per site; it diverges because we sent the Fermi energy to infinity when deriving the free energy.
Thus, the temperature-dependence of the free energy density in the normal state is
\beg
f_\mathrm{n}= -\frac{\pi^2 \nu_0 T^2}{3}- \nu_0g^2 \mbox{Re}[\psi(ia)]-\nu_0g^2\ln T.
\en
The corresponding specific heat (heat capacity per site)   is
\beg
C_\mathrm{n}=-T\frac{d^2\! f}{ dT^2}=\gamma_0 T\left[ 1+ \lam h\left(\frac{\Omega}{2\pi T}\right)\right],
\label{cn}
\en
where
\beg
\gamma_0=\frac{2\pi^2 \nu_0}{3},
\label{nonint_gam}
\en
 is the specific heat coefficient of free electrons, $\lam=\frac{g^2}{\Omega^2}$, and
\beg
h(x)=-6x^2-12 x^3 \mbox{Im}[\psi'(i x)]-6x^4\mbox{Re}[\psi''(ix)].
\en
The same expression~\re{cn}   obtains by a different method~\cite{grimvall}, which provides an independent check on the spin chain representation of the free energy.

We show three representative plots of $C_\mathrm{n}(T)$ in Fig.~\ref{c}.  Notice that, for instance, for $\lam=4.5$ the specific heat is negative in an interval of temperatures from $T_-\approx 0.4\Omega$ to $T_+\approx 0.7\Omega$. The superconducting transition temperature for large $\lam$ is $T_c\approx 0.18 \sqrt{\lam}\Omega$~\cite{allendynes},
which for $\lam=4.5$ is $T_c\approx 0.4\Omega$. Hence  the quasiparticle heat capacity is negative above $T_c$.

Let us determine the value of $\lam$ at which the specific heat  becomes negative for the first time. The function $h(x)$ in \eref{cn} has a single minimum $h_{\min}=-0.2709$ at $x_{\min}=0.3273$. The normal state specific heat $C_\mathrm{n}$ becomes negative for $\lam > - h_{\min}^{-1}\equiv \lam_*\approx 3.69$. The normal-superconductor transition for $\lam=\lam_*$ occurs at $T_c\approx 0.35 \Omega$. The minimum of the function $\frac{C_\mathrm{n}(T)}{ C_\mathrm{FL}}$, where $C_\mathrm{FL}=\gamma_0 T(1+\lam)$ is at $T_{\min}= (2\pi x_{\min})^{-1} \Omega\approx 0.49 \Omega$. At this temperature the specific heat is always negative for $\lam > \lam_*$. 

Therefore, for any $\lam>\lam_*$ the quasiparticle specific heat is negative  in an interval of temperatures $(T_-, T_+)$, where $T_+ > T_c$,
\beg
\mbox{$C_\mathrm{n}(T) < 0$ for $\lam > \lam_*\approx 3.69$ and $T_- < T<T_+.$}
\en
The length of this interval starts  from zero at $\lam=\lam^*$ and grows monotonically   with $\lam$. At first, both $T_+$ and $T_-$ are above $T_c$ until $T_-$ falls below  it. At large $\lam$
\beg
 T_-\approx 0.31\Omega,\quad T_+\approx 0.38\sqrt{\lam}\Omega,
\en
where we took the large $\lam$ asymptote of $T_+$ from \eref{tplas} below   and for $T_-$ we obtained it from $h(x_0)=0$. The numerical solution is $x_0\approx 0.5100$ and therefore $T_-=(2\pi x_0)^{-1}\Omega \approx 0.31\Omega$
for $\lam \to\infty$.

In the strong coupling limit $\lam\to\infty$ and $\Omega\to 0$, so that $g^2=\lam\Omega^2=\mbox{fixed}$.  Then, $T\gg\Omega$ and with the help of the series expansion for the digamma function, we find that \eref{cn} becomes
\begin{align}
\label{clargelam} C_\mathrm{n}=\gamma_0 T\left[1- \left(\frac{T_+}{T}\right)^2\right],\\ 
\label{tplas} T_+=g\sqrt{\frac{3}{2\pi^2}} \approx 0.38g.
\end{align}
This is negative for all temperatures below $T_+$ and again $T_+>T_c$. 

It is also instructive to  evaluate $C_\mathrm{n}$ in the strong coupling limit directly from \eref{fint}, where now $a=0$. We have
\beg
\frac{f_\mathrm{int}}{\nu_0 g^2}=\sum_{l=1}^{l_0} \frac{1}{l}=\int\limits_{2\pi T}^\Lambda \frac{d\omega_l}{\omega_l}+\gamma.
\label{fintst}
\en
Here we introduced a cutoff as discussed below \eref{digamma}. Combining this with $f_0$ in \eref{nif}, we obtain
\beg
f_\mathrm{n}= -\frac{\pi^2 \nu_0 T^2}{3}-\nu_0g^2\ln T.
\label{nsf_strong}
\en
The normal state entropy $S_\mathrm{n}$ and  specific heat $C_\mathrm{n}$ therefore are
\begin{align}
\label{Sn} S_\mathrm{n}= -\frac{ d f}{ d T}=\gamma_0 T + \frac{ \nu_0 g^2}{T},\\
\label{clargelam1} C_\mathrm{n}=-T\frac{ d^2\! f}{ d T^2}=\gamma_0 T - \frac{ \nu_0 g^2}{T}.
\end{align}
This $C_\mathrm{n}$ coincides with \eref{clargelam}.

Recall that $f$ is the contribution of the fermionic quasiparticles to the total free energy, which is $f$ plus the free energy of  noninteracting thermal phonons.  The combined specific heat  of quasiparticles   and  phonons   is positive.   For example, the specific heat of Einstein phonons in 3D at $T\gg\Omega$ is $C_E=3 n_i$. Assuming the number density of ions $n_i$ is the same as that of electrons, $C_E= 2 \nu_0 \eps_F $.
This is much larger in magnitude than the minimum   $C_\mathrm{n}\sim - \nu_0g $  above $T_c$ [see \eref{clargelam}], since   $\eps_F\gg g$. 
However, looking back at the derivation of the Eliashberg free energy in Ref.~\cite{spinchain}, we observe that the partition function of the   system is of the from $\mathcal{Z}=\mathcal{Z}_s \mathcal{Z}_{p}$. Here $\mathcal{Z}_s$ is the partition function of our classical spin chain or, equivalently, of the fermionic degrees of freedom and $\mathcal{Z}_{p}$ is the partition function of noninteracting phonons. Thus phonons and quasiparticles are two decoupled subsystems in the Migdal-Eliashberg theory as  true quasiparticles must be. Both subsystems should have positive heat capacities or the system is thermodynamically unstable~\cite{landau}.

We conclude that the quasiparticle picture  breaks down for large electron-phonon coupling $\lam$ together with the Migdal-Eliashberg theory  based on it. This result is independent of the model electron-phonon Hamiltonian, since at strong coupling the free energy functional  always converges to the spin chain Hamiltonian~\re{schol3} as  discussed in Sec.~\ref{strong_subsec}. The critical value of $\lam$ where the Eliashberg stationary point ceases to be the global minimum must be in any case no larger than   $\lam_*$ at which the quasiparticle specific heat turns negative. We expect the precise values of $\lam_c$ and $\lam_*$ to depend  only weakly on the underlying model, because $\lam_*\approx 3.69$ we obtained for the Holstein model is already quite deep in the strong coupling regime where all models converge.  There are reportedly~\cite{carbotte} materials (Pb$_{0.5}$Bi$_{0.5}$) with $\lam\approx 3.0$ that are well described by the Migdal-Eliashberg theory. Therefore
we expect
\beg
3.0\le \lam_c\le 3.7.
\label{range_lamc}
\en

\subsection{Specific heat and entropy in the superconducting state}
\label{sp_heat_sc_sec}

In stark contrast to the normal state, thermodynamics of the superconducting state is free of pathologies. The specific heat is positive
at any coupling strength and the entropy vanishes when $T\to0$ as it should. To show this, it is sufficient to analyze the worst case scenario  $\lam=\infty$.  In this limit, we are able to determine  thermodynamic properties at low temperatures and temperatures just below $T_c$ essentially analytically, while
 computing them  for general $\lam$
would require substantial numerical work. Since  the strong coupling limit of the Migdal-Eliashberg theory is model-independent, our results apply equally well to the Holstein Hamiltonian~\re{holsteinH} and the general electron-phonon  model~\re{frol1}.

Consider temperatures near $T_c$. The jump in the specific heat at $T_c$ for $\lam=\infty$ is~\cite{carbotte}
\beg
\Delta C=C_\mathrm{sc}-C_\mathrm{n}\approx 19.9\,\gamma_0 T_c.
\en
Setting $T_c\approx 0.18g$ in \eref{clargelam1}, we determine  the normal state specific heat at $T=T_c$,
 \beg
 C_\mathrm{n}(T_c)\approx -3.7 \gamma_0 T_c.
 \en
 Therefore the specific heat in the superconducting state at $T=T_c$ is
\beg
C_\mathrm{sc}(T_c)\approx 16.2\,\gamma_0 T_c.
\en
We see that the specific heat is positive just below $T_c$. 

Now let as evaluate the entropy and specific heat at low temperatures.
At the global minimum the spins $\bm S_n$ are coplanar~\cite{spinchain}. Choosing the $x$-axis so that $S_n^y=0$, we have
\beg
S_n^z=\cos\theta_n,\quad S_n^x=\sin\theta_n,
\en
where $\theta_n$ is the angle the spin makes with the $z$-axis. Expressing the spin chain Hamiltonian~\re{schol3} in terms of $\theta_n$, we obtain the free energy in the form
\beg
\begin{split}
f_\mathrm{sc}=&-2\pi\nu_0 T\sum_n\omega_n  \cos\theta_n\\
&-\pi^2\nu_0 T^2 g^2\sum_{nm} \frac{ \cos(\theta_n-\theta_m)-1}{(\omega_n-\omega_m)^2+\Omega^2}.
\end{split}
\label{ssf}
\en
The stationary point equation, $\partial f_\mathrm{sc}/\partial \theta_n=0$, for $f_\mathrm{sc}$ is
\beg
\omega_n\sin\theta_n= \pi T g^2\sum_{m} \frac{\sin(\theta_m-\theta_n)}{(\omega_n-\omega_m)^2+\Omega^2}.
\label{gapeqn}
\en
This is nothing but the Eliashberg gap equation  written in terms of $\theta_n$~\cite{spinchain}. 

The relationship between the gap function  $\Delta(i\omega_n)\equiv\Delta_n$ and $\theta_n$ is
\beg
 \cos\theta_n=\frac{\omega_n}{\sqrt{\omega_n^2+\Delta_n^2}},\quad  \sin\theta_n=\frac{\Delta_n}{\sqrt{\omega_n^2+\Delta_n^2}}.
 \label{theta_delta2}
 \en
The gap equation also follows from  \eref{Phieq} after we substitute
\beg
\Phi_n=\Delta_n Z_n,\quad\omega_n+\Sigma_n=\omega_n Z_n,
\label{varchange}
\en
 and
express $Z_n$ in terms of $\Delta_n$ from \eref{Sigmaeq}. Eliashberg equations~\re{elifamiliar3} become
\begin{subequations}
\begin{eqnarray}
\label{gapeq} \omega_n\Delta_n=\pi T\sum_m \lambda_{nm}\frac{ \omega_n  \Delta_m - \Delta_n  \omega_m}{\sqrt{\omega_m^2+|\Delta_m|^2}},\\
\label{Z} Z_n=1+\frac{\pi T}{\omega_n}\sum_m \lambda_{nm}\frac{\omega_m}{\sqrt{\omega_m^2+|\Delta_m|^2}}.
\end{eqnarray}
\end{subequations}
 The replacement~\re{theta_delta2} turns \eref{gapeq} into \eref{gapeqn}.

It is helpful to introduce the \textit{condensation energy}
\beg
\begin{split}
\delta f = & f_\mathrm{sc} - f_\mathrm{n}=-2\pi\nu_0 T\sum_n (\omega_n \cos\theta_n-|\omega_n|)\\
&- \frac{\nu_0   g^2}{4}\sum_{n\ne m} \frac{ \cos(\theta_n-\theta_m)-\mbox{sgn} (\omega_n\omega_m)}{(n-m)^2 },
\end{split}
\label{cond}
\en
where we took the strong coupling limit $\Omega\to 0$ and used $\omega_n=\pi T (2n+1)$. A helpful property of this expression is that
all sums in it converge as long as $\sum_{n>0} n\theta_n^2 <\infty$~\cite{michael} unlike in \esref{nsf} and \re{ssf}. This is important because, for example, it is due to the divergence of the double sum in \eref{nsf} that we gained the $\ln T$ term in \eref{nsf_strong}. Had this  sum
converged,  it would contribute only a temperature independent constant  with no effect on the entropy and specific heat.   Since $\Delta_n\to 0$ as $n\to\infty$,    \eref{theta_delta2} implies $\theta_n=o(n^{-1})$
and therefore $\sum_{n>0} n\theta_n^2 <\infty$.

 In the strong coupling limit,  differentiation of the condensation energy with respect to $T$ simplifies considerably, since the interaction
 term in \eref{cond} has no explicit temperature dependence -- its only dependence on $T$ is through $\theta_n$. We need $\delta f$ at the stationary points and because at these points $\partial[\delta f]/\partial \theta_n=0$, we have
 \beg
 \frac{d  [\delta f]}{ dT}=  \frac{\partial [\delta f]}{\partial T}+\sum_n  \frac{\partial [\delta f]}{\partial \theta_n} \frac{\partial \theta_n}{\partial T}=\frac{\partial [\delta f]}{\partial T}.
 \en
 Applying this formula to \eref{cond}, we find
 \beg
 \frac{d  [\delta f]}{ dT}= -8\pi\nu_0 \sum_{n=0}^\infty \left(\frac{\omega_n^2}{\sqrt{\omega_n^2+\Delta_n^2}}-\omega_n\right),
 \label{mats_sum}
 \en
 where $\Delta_n$ is the solution of the gap equation~\re{gapeqn}. We calculate this Matsubara sum in Appendix~\ref{matsubara}. Notably, we obtain an interesting identity along the way,
 \beg
 \int_0^\infty\!\!\! d\omega \left(\omega-\frac{\omega^2}{\sqrt{\omega^2+{\Delta }^2(i\omega)}}\right)=\frac{g^2}{4}.
 \label{iden124}
\en
Here $\Delta(i\omega)$ is  the Eliashberg gap  function on the Matsubara axis at zero temperature in the strong coupling limit.

 The end result for the entropy $S_\mathrm{sc}$ and specific heat $C_\mathrm{sc}$ in the superconducting state at low $T$ and $\lam=\infty$ is (see Appendix~\ref{matsubara})
\begin{align}
 S_\mathrm{sc}\approx 17.84 \nu_0 \frac{E_1}{T} e^{-E_1/T},\quad E_1\approx 1.16g,\\
 C_\mathrm{sc}\approx 17.84 \nu_0 \left(\frac{E_1}{T}\right)^2 e^{-E_1/T}.
 \end{align}
The specific heat is positive and the entropy vanishes when $T\to 0$ as it should.

Therefore, 
there are no apparent pathologies in the thermodynamics of the Migdal-Eliashberg superconducting state. Of course, this does not prove  this  state  is necessarily the global minimum of the free energy below $T_c$ and we will later see that in fact it is not at least
in some range of temperatures.

\subsection{Quasiparticle lifetime: normal state}
\label{selfsect}

It is natural to confirm the breakdown of the quasiparticle picture    by analyzing  quasiparticle lifetimes   at large $\lam$. 
Consider the normal and anomalous thermal Green's functions defined as
\begin{align}
\GG_{\pp} (\tau-\tau')=- \langle T_\tau c_{\pp \sigma}(\tau)  c^\dagger_{\pp\sigma}(\tau')\rangle,\label{GG}\\ 
\FF_\pp (\tau-\tau')= \langle T_\tau c_{-\pp\dn}(\tau)  c_{\pp\up}(\tau')\label{FF}\rangle.
\end{align}
In the Migdal-Eliashberg  theory,    these Green's functions are in the Matsubara frequency domain (see, e.g., Ref.~\cite{spinchain})
\begin{align}
 \label{gsym}\GG_{\pp n}=-\frac{ i(\omega_n+\Sigma_{n} )  +\xi_\pp } {(\omega_n+\Sigma_{n} )^2 +| \Phi_{n} |^2 +  \xi_\pp^2  },\\
\label{fsym}\FF_{\pp n}=-\frac{ \Phi_{n}   } { (\omega_n+\Sigma_{n} )^2 +| \Phi_{n} |^2 +  \xi_\pp^2}.
\end{align}
In the normal state, $\Phi_n=0$ and therefore
\beg
\mathcal{G}_\pp(\omega_n)=\frac{1}{i\omega_n+i\Sigma_n -\xi_\pp}.
\en
Further, \esref{varchangeR}, \re{sgf} and \re{nsspins} imply
\beg
 \Sigma_n=\pi T  \sum_m \lam_{nm}  \mbox{sgn} (\omega_n).  
 \label{selfnm}
 \en
 Since in the strong coupling regime the characteristic phonon frequency $\Omega\to 0$, we take $T$ to be much greater than $\Omega$.
 Then, the $n=m$ term dominates  the summation and we obtain
 $\Sigma_n= \lam\pi T\mbox{sgn}(\omega_n)$ and therefore
\beg
\mathcal{G}_\pp(\omega_n)=\frac{1}{i\omega_n -\xi_\pp+i \lam\pi T\mbox{sgn}(\omega_n) }.
\en
It is straightforward to analytically continue this to the upper half-plane of complex $\omega$~\cite{omega} to get the retarded Green's function
\beg
G_\pp^R(\omega)=\frac{1}{\omega -\xi_\pp+i \lam\pi T }.
\label{GR}
\en
Observe that   the quasiparticle decay rate in the normal state
\beg
\Gamma_\mathrm{n}=\tau_\mathrm{n}^{-1}=\lam \pi T
\en
 is much larger than the temperature. The lifetime $\tau$ tends to zero as $\lam\to\infty$. This means that the fermionic quasiparticles are ill-defined   in agreement with our specific heat argument.  
  
 \subsection{Quasiparticle spectrum in the superconducting state}
 
Before we estimate quasiparticle lifetimes in the superconducting state at large $\lam$, we need to  know  the properties of the excitation spectrum in this regime. We  will see that   the low energy part of the quasiparticle  spectrum consists of narrow bands of width $g\lam^{-1/2}$. The gaps between the bands decrease with energy $E$ as $E^{-1}$ until the spectrum becomes continuous above $E_{ct}\sim g \lam^{1/2}$,  when the bandwidth is comparable to the gaps. This is consistent with the expectation of Fermi-liquid-like spectrum at energies of the order of $\eps_F$. Indeed,   Migdal's theorem~\cite{migdal} requires
\beg
x_M=\frac{\lam\Omega}{\eps_F}\ll1.
\label{mparameter}
\en 
Since $\lam= g^2/\Omega^2$, this  implies  $\eps_F\gg E_{ct}$.

 To determine the spectrum, we first obtain the leading large $\lam$ asymptotic behavior of $Z_n$ from \eref{Z}
 \beg
 Z_n= \lam \pi T(\omega_n^2+\Delta_n^2)^{-1/2}.
 \label{strongZ}
 \en
  Since $\Delta_n$  remains finite the limit $\lam\to\infty$~\cite{combescot,mars_strong}, $Z_n$ diverges at any finite $\omega_n$. Assuming $\xi_\pp$ is also finite and  performing the variable change~\re{varchange} in \eref{gsym}, we find
 \beg
 \GG_{\pp n}=\frac{ -i\omega_n  } {\lam\pi T\sqrt{\Delta_n^2+\omega_n^2}  },
 \en
 where we substituted $Z_n$ from \eref{strongZ}. Analytic continuation to the upper half plane~\cite{omega} gives
\beg
 G^R_{\pp}(\omega)=\frac{ -\omega } {\lam\pi T\sqrt{\Delta^2(\omega)-\omega^2}  }.
 \label{grs}
 \en
 Recall the   Lehmann representation for the retarded Green's function~\cite{mahan}:
 \beg
  G^R_{\pp}(\omega)=\sum_k \frac{| \langle k| c_\pp|0\rangle|^2}{\omega-E_k+i0^+}+ \sum_k \frac{| \langle 0| c_\pp|k\rangle|^2}{\omega+E_k+i0^+}.
  \label{lehman}
 \en
  Here $|k\rangle$ are the eigenstates of the electron-phonon Hamiltonian,  $E_k$ are single electron excitation energies (energy
  differences between  eigenstates with  $N_\mathrm{e}\pm1$ electrons and the ground state with $N_\mathrm{e}$ electrons), and  we set $T=0$.
  
 Comparing \esref{grs} and \re{lehman}, we conclude that
 \beg
 \frac{ \omega } { \sqrt{\Delta^2(\omega)-\omega^2}} = \sum_k\left( \frac{ P_k}{\omega-E_k}+   
 \frac{ P_k}{\omega +E_k}\right)\!,
 \label{compare}
 \en
 where we absorbed $i0^+$ into $\omega$ and
 \beg
 P_k=\pi \lim\limits_{T\to0} \lim\limits_{\lam\to\infty} \left(\lam T | \langle k| c_\pp|0\rangle|^2\right).
 \en
 $P_k$ must be finite and well-defined, because $\Delta(\omega)$ is finite and well-defined in this limit. Residues  at $\omega=\pm E_k$ are equal by particle-hole symmetry [see the discussion above \eref{feH1}]. The limits $\lam\to\infty$ and $T\to0$ commute for the gap function -- one obtains the same $\Delta(\omega)$ no matter in which order these limits are taken~\cite{combescot}. However, they do not   commute in general. For example, we saw in the previous paper~\cite{spinchain},   that there are solutions of the Eliashberg equations that are present for one order of limits and absent for the other. We always take the limit $\lam\to\infty$ first. Note also that the density of quasiparticle states at any $\lam$ is
 \beg
 \frac{\nu(\omega)}{\nu_0}=  \mbox{Im}\left[ \frac{\omega}{\sqrt{\Delta^2(\omega)-\omega^2}}\right],
 \label{dos}
 \en
 which we derive by integrating \eref{gsym} over $\xi_\pp$. 
 
 Equation~\re{compare} has several remarkable consequences. Consider real values of $\omega$. First, because the right hand side is real, $\Delta(\omega)$ must also be real   except for a discrete set of points (zeros of the right hand side) where $\mbox{Im }\Delta(\omega)$ must be infinite. In other words, $\mbox{Im }\Delta(\omega)$ is a sum of delta functions. Moreover, $|\Delta(\omega)|\ge |\omega|$ for the same reason. Second, excitation energies
 $\pm E_k$ are solutions of the equation
 \beg
 \Delta(\omega)=\pm\omega.
 \label{exc}
 \en
 The roots of this equation are necessarily doubly degenerate, since the right hand side of \eref{compare} has poles rather than branching points at these values of $\omega$. This also follows from   $|\Delta(\omega)|\ge |\omega|$ as this inequality implies that at $\omega=\pm E_k$ one of the lines $\pm\omega$ is tangent to   $\Delta(\omega)$. 
 
 Most important for our purpose is the observation that solutions of \eref{exc} form a discrete set and therefore the low energy quasiparticle  spectrum is discrete. Indeed, two analytic functions cannot coincide on an interval without being identically equal. Since \eref{exc} does not hold for all
 $\omega$, it can hold only at a discrete set of points $\pm E_k$, where $k=1,2,\dots$. These corollaries of \eref{compare}  reproduce and confirm the results of a
 more thorough study of the quasiparticle spectrum in the strong coupling limit by Combescot~\cite{combescot}. Since $g$ is the only energy scale left in this limit, $E_k/g$ are numbers of order one. In particular, Combescot finds, $E_1=1.16g$ and $E_2=3.04g$, while for large $k$
 \beg
 E_k=\pi g\sqrt{k}.
 \label{largek}
 \en
 Levels $E_k$ are macroscopically degenerate with the degree of the degeneracy controlled by the residue $P_k$ in \eref{compare}.  It is interesting to note here that the excitation  spectrum  of the BCS model  in the strong coupling limit is a discrete set of macroscopically degenerate levels
 as well~\cite{baytin}.  
 
  At finite $\lam$   levels $E_k$ split into energy bands.  We show in Appendix~\ref{correction_sec} that the width of these bands is approximately $\Omega=g \lam^{-1/2}$.   It follows from \eref{largek}   that the gaps between bands decrease as $\frac{\pi g}{2\sqrt{k}}$ with the band number $k$. The spectrum becomes continuous when the bandwidth becomes equal to the gap, i.e., for
  \beg
  E\ge E_{ct}=\frac{\pi^2 g \sqrt{\lam}}{2}.
  \en
And indeed we expect continuous spectrum   at energies of the order of the Fermi energy, much larger than typical energies associated with superconductivity. At such energies the system must be a Fermi liquid. For sufficiently  large $\omega_n$, 
$|\Phi_n|=Z_n\Delta_n$ is much smaller than $\omega_n+\Sigma_n=Z_n\omega_n$
 in the normal Green's function $\GG_{\pp n}$ given by \eref{gsym}.  Neglecting $|\Phi_n|$, we obtain the normal state Green's function~\re{GR} and recover  Fermi liquid dispersion $\xi_\pp$.
 One has to be careful here because, while $\Delta_n$ is of order
 $g$ at large $\lam$ and quickly decreases for $\omega_n>g$, the gap function $\Delta(\omega)$ along the real frequency axis does not necessarily behave in the same way.  Along the real axis $\Delta(\omega)$  should in fact decrease substantially only at energies where the spectrum becomes continuous, i.e., at an energy scale $E_{ct}\gg g$. Nevertheless, the smallness of the Migdal's parameter $x_M$ guarantees that  the Fermi energy is even larger as seen from
 \eref{mparameter}.
  
  \subsection{Quasiparticle lifetime: superconducting  state}
  
  As with thermodynamic properties, the situation with quasiparticle decay  predicted by the Migdal-Eliashberg theory for  the superconducting state at strong coupling is in some sense opposite to that in the normal state.  Consider $\lam=\infty$ first.
   Equation~\re{compare} shows that the density of states at $T=0$  is a sum of delta-functions. The width of  quasiparticle peaks at $E_k$ is zero and the lifetime is therefore infinite. At $T\ne0$ thermally activated transitions between quasiparticle
   energy levels $E_k$ occur. However, their rate  is exponentially small at low temperature. Direct transitions with an absorption or emission of a phonon are prohibited because the phonon energy $\Omega/g\to0$, while the spacing between $E_k$ is of order $g$. Instead,  phonons provide a thermal bath for electrons inducing transitions via thermal noise. Since $T_c\approx 0.18 g$, at temperatures well below $T_c$ the thermal energy $T$ is much smaller than the typical spacing between the levels. 
   
   Consider several examples of scattering processes. A quasiparticle on level $E_1$ can interact with and break a Cooper pair resulting in three $E_1$ quasiparticles,   $E_1 \to 3E_1$ and $2E_1/\Omega$ phonons. It can also absorb $N_\mathrm{ph}=(E_2-E_1)/\Omega$ phonons and make a transition to level $E_2$, i.e., $E_1\to E_2$. An $E_2$ particle can   emit phonons and turn into an $E_1$ quasiparticle ($E_2\to E_1$) or it can  break a Cooper pair along the way resulting in three $E_1$ quasiparticles ($E_2\to 3E_1$). Since the electron-phonon interaction  [see, e.g., \esref{holsteinH} and \re{frol1}] can change the phonon number only by one at a time, all these processes have to go through multiple virtual states, e.g., $|E_2\rangle \to |E_1,  1\rangle \to |E_2, 2 \rangle \to\dots |E_1, N_\mathrm{ph} \rangle$, where we indicated the number of phonons at the second position in the ket vector. Intermediate states here   are virtual    and there are transitions with energy barriers of order $g$ in  any such process.  Then, according to Kramer's rate 
   theory~\cite{kramers1,kramers2,kramers3}
 the quasiparticle decay (tunneling) rate for $T\ll g$ is
 \beg
 \Gamma_\mathrm{sc}  =c_1 g e^{-c_2 g/T},
 \label{intraband}
 \en
 where $c_1$ and $c_2$ are numerical coefficients of order one.  
  
 Now let $\lam$ be large but finite. We saw  above that  for such $\lam$ the energy level $E_k$ broadens into a  band of width $\Omega$. It is natural to interpret 
 this bandwidth as the uncertainty in the quasiparticle energy. Its inverse is then the quasiparticle lifetime  and therefore the quasiparticle decay rate  at low energies  is
 \beg
  \Gamma_\mathrm{sc}^\mathrm{in} =\Omega= g \lam^{-1/2}.
  \label{gammalam}
  \en
 Equation~\re{intraband} gives the rate of thermally activated  transitions between different energy bands and \eref{gammalam} -- the rate of transitions within a band.  The total decay rate is the sum of the two
 \beg
 \Gamma_\mathrm{sc}^\mathrm{ tot} =c_1 g e^{-c_2 g/T}+  g \lam^{-1/2}.
 \en
 Since $g/T_c\approx 5.5$, the second term dominates for all but extremely large $\lam$. In any case,  quasiparticle lifetime is very large at low energies. This again shows that there is a certain robustness, rigidity to the superconducting state. This state is not as manifestly unstable as the normal state. 
 
\section{Qualitative picture of the breakdown}
\label{qualitative_sect}

Let us develop a more intuitive  understanding of the breakdown of the Migdal-Eliashberg theory. We seek to explain vanishing quasiparticle lifetime and negative specific heat in the normal state at strong coupling and why the superconducting state is free of such pathologies. Of these two negative specific heat is especially important as it defines a value $\lam_*\approx 3.69$ of the electron-phonon coupling above which the theory becomes invalid.
 We will see that the diverging   quasiparticle decay rate   is due to scattering of electrons from thermal fluctuations of  static   displacements of the ions (classical phonons), which have a natural interpretation as a  disorder potential. The superconducting state is not affected by static disorder by Anderson's theorem.

The mechanics behind negative quasiparticle heat capacity is more sophisticated. We will see that this thermodynamic instability  is driven by electrons near the Fermi surface interacting via quantum phonons (quantum fluctuations of the lattice). At $\lam=\lam_c$ the quasiparticle band structure changes abruptly   and these electrons and quantum phonons lower their energy by forming new bound states. The Migdal-Eliashberg treatment does not capture the emergence of these new fermionic quasiparticles, but signals it via negative specific heat.
     
\subsection{Quasiparticle decay rate}
\label{quasidecay}

We found above that the quasiparticle decay rate in the normal state at strong electron-phonon coupling $\lam$ is
\beg
\Gamma_\mathrm{n}=\lam \pi T.
\label{norm_rate1}
\en
   There are two ways to interpret this result. First, it is important to realize that  it is entirely due to electrons scattering  from static displacements of the  ions, which act as  nonmagnetic impurities. 

Consider the electron-phonon interaction term in the Holstein Hamiltonian~\re{holsteinH}
\beg
H_\mathrm{el-ph}=\sum_{\bm i} (\alpha x_\ii) n_{\bm i}.
\label{holsteinHint}
\en
 At strong coupling any finite temperature $T$ is much larger than the frequency $\Omega\to0$ of lattice oscillators. The oscillators are   highly excited and  therefore essentially classical. Their momenta $p_\ii$   decouple and integrate out in the partition function. We are left with  their coordinates $x_\ii$ which are classical variables independent of the imaginary time $\tau$ -- \textit{classical phonons}.  Therefore, $ \alpha x_\ii\equiv V_\ii$ in \eref{holsteinHint}  is equivalent to a single-particle potential for the electrons. The problem is    that of electrons
moving in a random (due to thermal fluctuations of $x_\ii$) potential $V_\ii$. The potential comes at an elastic energy cost $\sum_\ii K x_\ii^2/2$, where $K$ is the renormalized spring constant of the oscillators. The classical variable $x_\ii$ coincides
with its zeroth Matsubara component $x_\ii(0)$, i.e., with the imaginary time average $\bar x_\ii$
\beg
 x_\ii = \bar x_\ii\equiv x_\ii(0)= \frac{1}{\beta}\int_0^{\beta}\!\!\!  d\tau\, x_\ii(\tau),\quad \beta=\frac{1}{T},
\en
since $x_\ii(\tau)=x_\ii$ is $\tau$-independent.
For this reason, we also refer to classical $x_\ii$ as  static displacements or  classical phonons and use the notation   $\bar x_\ii$ for them instead of $x_\ii$ from now on to avoid confusion with the general quantum case. 
Note also that nonzero Matsubara components  account for quantum fluctuations of ionic positions.
 
Quasiparticle decay rate due to nonmagnetic impurities is~\cite{agd,lee}
\beg
\Gamma_\mathrm{imp}=\pi \nu_0 V^2.
\en
The quantity $V$  is the average strength of the disorder potential   defined through
\beg
\sum_\ii \langle V_\ii(\jj) V_\ii(\jj') \rangle= V^2 \delta_{\jj\jj'},
\label{imp_pot}
\en
where $V_\ii(\jj)$ is the potential at site $\jj$ produced by the impurity at $\ii$.
 In our case, $V_\ii(\jj)=\alpha \bar x_\ii \delta_{\ii\jj}$ and the average in \eref{imp_pot} is the thermal average. We obtain
\beg
\Gamma_\mathrm{imp}=\pi \nu_0 \alpha^2 \langle  \bar x_\ii^2\rangle_T,
\en
By equipartition theorem for a classical harmonic oscillator, $K \langle \bar x_\ii^2\rangle_T=T$. Using this and the definition of $\lam$ in \eref{lam}, $\lam=\nu_0\alpha^2/K$, we find that $\Gamma_\mathrm{imp}=\lambda\pi T =\Gamma_\mathrm{n}.$ Therefore, to the leading order in the electron-phonon coupling $\lam$ the quasiparticle decay rate in the normal state   is due
to electrons  scattering from static displacements of the ions, or, in other words, from classical, zero Matsubara frequency   phonons.   Recall also that we previously obtained $\Gamma_\mathrm{n}=\lambda\pi T$   from the $n=m$ term in \eref{selfnm}, i.e.,   from the zero phonon frequency part of the self-energy.

Within this framework it is also easy to explain why  the quasiparticle decay rate  in the superconducting state remains negligible when at the same time it diverges in the normal state as $\lam\to\infty$. The answer is that, as we know from Anderson's theorem~\cite{dirty}, nonmagnetic disorder  does not affect superconducting properties in conventional superconductors. This also explains the reason behind the cancellation of   zero Matsubara frequency phonon ($\omega_n-\omega_m=0$ term)  from the Eliashberg gap equation~\re{gapeq} and the free energy~\re{schol}. Note that it is important here that the thermal averages $\langle x_\ii\rangle=0.$
The case of a regular pattern of nonzero $\langle x_\ii\rangle$ is not covered by Anderson's theorem.

Another interpretation of \eref{norm_rate1} is as a rate of phonon emission and absorption in the limit of zero phonon frequency $\Omega$. By Fermi's golden rule this rate is
\beg
\begin{split}
\Gamma_\mathrm{ph}=2\pi {\tilde g}^2  \Bigl\{& n_B(\Omega)\left[ 1-n_F(\Omega)\right]  \Bigr.\\
 & + \Bigl.\left[n_B(\Omega)+1\right] \left[ 1-n_F(-\Omega)\right]  \Bigr\}\nu_0.
\end{split}
\label{gammaph}
\en
Here $n_B$ and $n_F$ are Bose and Fermi distributions and $\tilde g= \alpha/\sqrt{2 M\Omega}$ is the electron-phonon interaction strength, which we obtain from \eref{frol1} by setting $\alpha_\qq=\alpha$ and $\omega_0(\qq)=\Omega$.
The first term in \eref{gammaph} corresponds to a fermion at the Fermi level $\eps_F=0$ absorbing a phonon of energy $\Omega$ and making a transition to the level $\Omega$ as long as that level is empty. The second term describes spontaneous plus simulated emission of a phonon by an electron at the Fermi level. Using $1-n_F(-\Omega)=n_F(\Omega)$ and the definitions of $g^2$ and $\lam$ in \esref{Holstein_lambda} and \re{lam}, we obtain the standard expression for inverse electron lifetime specialized to the case of  Einstein phonons~\cite{mahan}
\beg
\Gamma_\mathrm{ph}=\pi  \lam\Omega\{ n_B(\Omega)+n_F(\Omega)\}.
\en
When $\Omega/T\to0$, the distributions $n_B(\Omega)\to T/\Omega$ and $n_F(\Omega)\to 1/2$. Therefore, in this limit
$\Gamma_\mathrm{ph}=\lam\pi T=\Gamma_\mathrm{n}$ as claimed. 

\subsection{Negative specific heat}

\begin{figure}[htb]
\begin{center}
\setlength{\unitlength}{10cm}
\begin{picture}(0.85, 0.45)(0,0)
   \put(0,0){\resizebox{0.8\unitlength}{!}{\includegraphics{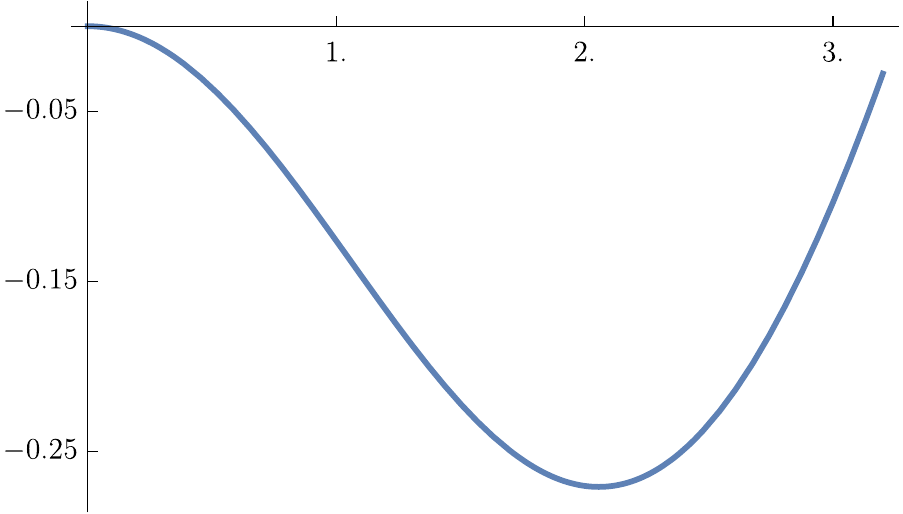}}}
    \put(0.81, 0.403){\makebox(0,0)[b]{ $\displaystyle \frac{\Omega}{T}$}}
   \put(0.12, 0.055){ {\makebox(0,0)[t]{  $ \displaystyle\frac{C_\mathrm{int} }{\lam  C_0 }$}}}
\end{picture}
\end{center}
\caption{Interaction contribution to the quasiparticle specific heat, $C_\mathrm{int}$ predicted by the Migdal-Eliashberg theory in 
units of $\lam C_0$, where $C_0$
 is the specific heat of free fermions, $\lam$ is the   electron-phonon coupling, and $\Omega$ is the natural frequency of the Einstein phonons. Note that $C_\mathrm{int}<0$ at any $\lam$ for all temperatures $T >\Omega/3$. }
\label{cint}
\end{figure}

 We derived the normal state specific heat $C_n$ within the Migdal-Eliashberg theory in Sec.~\ref{normal_heat_sec}, see \eref{cn}. The contribution of the  electron-electron interaction  is
\beg
\frac{C_\mathrm{int}}{C_0} = \lam h\left(\frac{\Omega}{2\pi T}\right),
\label{cnint}
\en
where $C_0=\gamma_0   T$ is the noninteracting part and the total specific heat is $C_\mathrm{n}=C_0+C_\mathrm{int}$. A plot of \eref{cnint} is shown in Fig.~\ref{cint}. We see that $C_\mathrm{int}$  is negative as long as   $T>\Omega/3$. It is also proportional to $\lam$, because the electron-electron interaction carries an overall factor of $g^2\propto \lam$, which corresponds to two electron-phonon vertices. Therefore,  $|C_\mathrm{int}|$  exceeds $C_0$ at any $T>\Omega/3$ for  large enough $\lam$, at which point
the total specific heat becomes negative.

 This instability  is driven    by \textit{quantum} 
 phonons, i.e.,  by quantum fluctuations of the ion displacements $x_\ii$.  Phonons that determine the quasiparticle  heat capacity are virtual and purely quantum because, as mentioned above, classical phonons sit at zero Matsubara frequency   and their contribution -- the $n=m$ term in \eref{schol3} --   cancels from the free energy.  For this reason,  this effect is   more subtle than   the divergence of the quasiparticle decay rate $\Gamma_\mathrm{n}$, which is entirely due to classical phonons.   Note also that unlike  negative specific heat, the linear growth of $\Gamma_\mathrm{n}$ does not provide a sharply defined value of  $\lam$ above which the Migdal-Eliashberg theory loses validity. 
  
 Consider   $\lam=\infty$ for simplicity. In the spin language, negative quasiparticle heat capacity comes from the hard   jump of   the $z$-component of spin   (Fig.~\ref{spinsfig}) combined with $\Omega=g\lam^{-1/2}=0$. It is these two factors that produce the divergent summation in \eref{fintst} and the problematic $\ln T$ term in the normal state free energy.   This term comes from interactions between  antiparallel   spins at $\omega_n>0$ and   $\omega_m<0$  at distances $\omega_l=|\omega_n-\omega_m|$ of order $2\pi T$ from each other, since  this contribution determines the lower limit of integration in \eref{fintst}. Therefore  virtual phonons with frequencies of the order of $2\pi T$ and  electrons with energies    of the same order  are responsible for the instability. In other words, the instability is due to interactions between electrons in a  window of order $2\pi T\ll \eps_F$ around the Fermi level mediated by quantum phonons.

  Recall
 that $S_n^z$ is proportional to an integral of the normal thermal Green's function~\re{gsym} over $\xi_\pp$.   Since the Fermi energy is by far the largest energy scale,  we integrate over $\xi_\pp$ from $-\infty$ to $+\infty$ with a constant density of states.  In the normal state, $\Phi_n=0$ and the integration gives $S_n^z=\mbox{sgn}(\omega_n+\Sigma_n)=\mbox{sgn}(\omega_n)$. In the superconducting state,  the same integration obtains
 \begin{subequations}
\begin{eqnarray}
 S_n^z&=&\frac{\omega_n + \Sigma_n}{\sqrt{ (\omega_n+\Sigma_n)^2+ |\Phi_n|^2}}=\frac{\omega_n}{\sqrt{\omega_n^2+\Delta_n^2}},\\
 S_n^x&=&\frac{\Delta_n}{\sqrt{\omega_n^2+\Delta_n^2}}.
 \end{eqnarray}
 \end{subequations}
 Below $T_c,$ spins acquire   $x$-components softening the jump in $S_n^z$. This  deviation of spins from the $z$-axis increases their ferromagnetic interaction energy resulting in a discontinuity in the specific heat, such that it becomes positive in the superconducting state as we found in Sec.~\ref{sp_heat_sc_sec}.  In this way, opening of the superconducting gap removes the instability.
 
 We saw in the previous subsection that classical ion displacements $\bar x_\ii$ provide a fluctuating single-particle potential
 $V_\ii=\alpha \bar x_\ii$ for the electrons.   Nonzero thermal averages of $\bar x_\ii$ mean a nonzero average  potential $V_\ii,$ which    modifies the electronic band structure. In particular, as we  discuss in more detail in Sec.~\ref{adiabatic}, it can open a gap $\Delta_P$ at the Fermi level via  the Peierls mechanism. This metal-insulator transition stabilizes the system   like  the opening of the superconducting gap. Indeed, suppose $\xi_\pp^2=\eta_\pp^2+\Delta_\pp^2$. Now the integration of \eref{gsym} 
 over $\eta_\pp$ from $-\infty$ to $+\infty$ in the normal state ($\Phi_n=0$) gives
 \beg
 S_n^z=\frac{\omega_n + \Sigma_n}{\sqrt{ (\omega_n+\Sigma_n)^2+ \Delta_P^2}}.
 \label{Snz_gap}
 \en
 We see that the band gap $\Delta_P$ plays a role similar to the anomalous average $|\Phi_n|$. Following the same steps as before~\cite{spinchain}  but for a gapped single-particle spectrum, we derived the spin chain representation for the free energy for this case. The part involving $S_n^z$ is the same as in \eref{schol3} but with $S_n^z$ from \eref{Snz_gap}.  In addition, there is
 an infinite range ferromagnetic $S_n^x S_m^x$ interaction. Stronger ferromagnetism suggests that  spectral gap opening precedes the superconducting transition in agreement with our finding that the specific heat becomes negative above the superconducting $T_c$.
 
  It is possible that a soft gap or a pseudogap may stabilize the electron-phonon system as well. However, we show in Sec.~\ref{adiabatic} that at least for  certain system  parameters a hard gap $\Delta_P$ (metal-insulator transition) is preferred.  In any case, a substantial  depression of the density of states near the Fermi energy at $\lam>\lam_c$ is necessary to remove the negative specific heat pathology. Other changes of the band structure, such as band narrowing etc.,  are insignificant  near $\lam_c$ given that the Fermi energy is still much larger than all other   energies.  Even though we discussed classical phonons separately for the sake of the argument, the effect of quantum phonons on the quasiparticle spectrum is equally important and inseparable from that of classical phonons.  
   
Now we are in a position to explain the breakdown of the Migdal-Eliashberg theory signaled by the negative  specific heat.
At $\lam_c$ the nature of fermionic quasiparticles changes abruptly. Electrons near the Fermi surface and quantum phonons lower their energy by  forming new bound states  -- new quasiparticles with gapped spectrum. This transition involves both quantum and classical phonons. Quantum phonons dress the electrons and classical phonons  facilitate the gap opening. Suppose  we prepare the system in the Migdal-Eliashberg normal state at $\lam>\lam_c$ and ``temperature'' $T$.  Here $T$ is a parameter rather than the true temperature as this state is not the true thermal equilibrium.  Next, we bring the system into contact   with  a thermal bath at temperature $T+\delta T>T$ and allow it to equilibrate.   Since there are new  quasiparticle states with lower energies available, some of the Migdal-Eliashberg quasiparticles transition into  these new states. The total energy  decreases as the system equilibrates, i.e., the  heat capacity is negative.

\section{New phase transition}
\label{new_order_sect}

We showed that the specific heat  of the Migdal-Eliashberg normal state is negative in a range of temperatures, $T_c< T< T_+$, at strong electron-phonon coupling, $\lam>\lam_*\approx 3.7$. Therefore, this state  is no longer the global minimum of the free energy. 
New order must emerge above certain $\lam_c$, such that   $3.0\le \lam_c\le 3.7$, see \eref{range_lamc}. Considerations of the preceding section suggest that    ion displacements acquire site-dependent averages $\langle  x_\ii \rangle$ breaking  lattice translational invariance in the emergent phase.  We also saw that $V_\ii=\alpha  x_\ii$ plays the role of a disorder potential. As  $\lam$ increases, the strength 
of the disorder $\alpha\propto \sqrt{\lam}$ increases with it. This again suggests   a metal-insulator transition. Indeed, we find  in Sec.~\ref{adiabatic} that   in the adiabatic limit at half filling the system is  an insulator for $\lam>\lam_c$. By continuity we expect  
this to persist at least to some extent into the non-adiabatic regime, see also Ref.~\cite{scalapino}. Another candidate for the new order is a Fermi liquid with broken lattice translational invariance. Whether the new global minimum is an insulator or such a Fermi liquid 
depends on  factors unimportant in standard Migdal-Eliashberg treatment, such as the filling fraction and lattice symmetry. We assume it is an insulator in this section for definiteness. 

\begin{figure}[htb]
\centerline{\includegraphics[width=0.9\columnwidth]{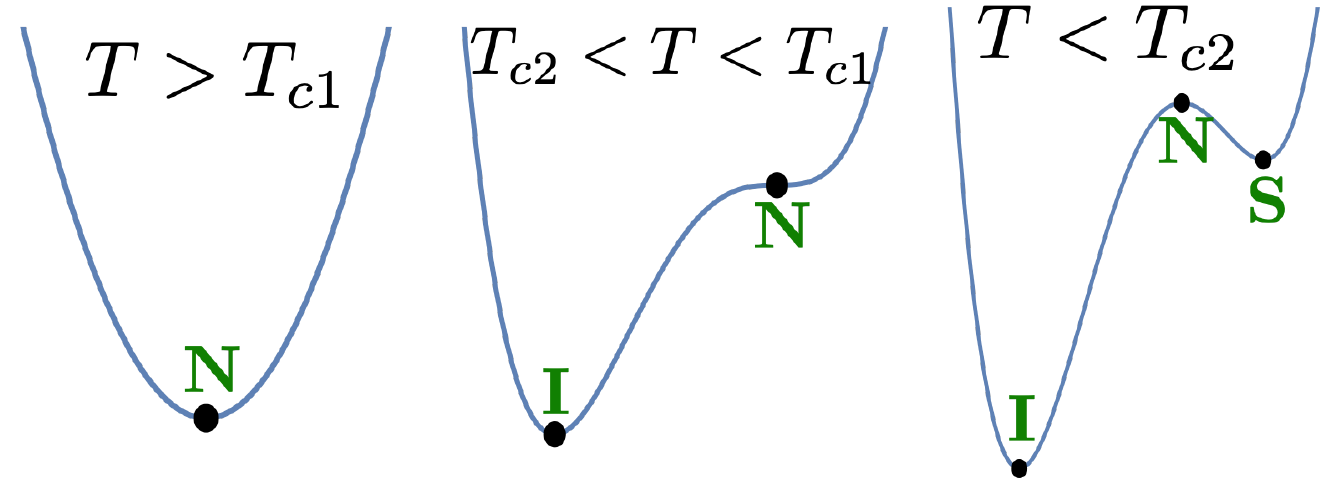}}
\caption{Schematic plot of the free energy of the electron-phonon system illustrating the emergence of a new global minimum    at strong coupling, $\lam>\lam_c$. This minimum is either an
insulator  or a Fermi liquid with broken  translational invariance; we take it to be an insulator for definiteness. At $T>T_{c1}$ the system is in the normal (N) state. At $T_{c1}$ insulating (I) order develops. As we decrease the temperature further, the superconducting (S) stationary point emerges at $T=T_{c2}$ as a local minimum or a saddle point.  At temperatures just below $T_{c2}$, this stationary point must be higher in energy than the insulator. }
\label{shape}
\end{figure}

When $\lam<\lam_c$, the system undergoes  a metal-superconductor transition at $T_c\equiv T_{c2}$ described by the Migdal-Eliashberg theory. At fixed $\lam>\lam_c$, the new phase transition occurs   at a certain critical temperature $T_{c1}>T_+>T_{c2}$. At very high temperatures the system is in the normal state (a classical gas of fermions and phonons). The superconducting stationary point develops below $T_{c2}$  as a local minimum or a saddle point, since the Eliashberg gap equation   has a nontrivial solution below $T_{c2}$ for all $\lam$. The superconducting state cannot be the global minimum just below $T_{c2}$, because it is close in energy to the normal state, while the  insulating state is already far, see Fig.~\ref{shape}.  Nevertheless, as we continue to lower the temperature, the superconductor can still prevail over the insulator via a first order phase transition.

One more consequence of the emergence of the new global minimum   is that there must be a first order phase transition
as a function of $\lam$ for certain  temperatures below $T_{c2}$.  We saw that there must be a range of temperatures below $T_{c2}$ where the system is
an insulator for $\lam>\lam_c$. As we decrease $\lam$ below $\lam_c$, the electron-phonon system switches from a well-formed insulating global minimum to a well-formed superconducting minimum. This is only possible through a first order phase transition.

\section{Comparison to other studies}
\label{compare_sect}

There are numerous publications  discussing the breakdown of the Migdal-Eliashberg theory at strong  coupling, see, e.g., Refs.~\cite{millis,roland,alexandrov,meyer,capone,scalapino,esterlis}. However, none of them demonstrate a true breakdown, i.e.,  show that the theory looses validity when the   coupling  $\lam$ exceeds
a certain finite value. Rather than  testing the validity of the Migdal-Eliashberg theory within its domain of applicability, most studies  rediscover the lattice instability pointed out by Migdal and Eliashberg~\cite{migdal,eli1st} or explore the post-instability physics to which the   theory no longer applies, see also the discussion in Introduction.

 In conventional electron-phonon models, such as the Fr\"olich and Holstein Hamiltonians,  electron-phonon interactions  renormalize the phonon frequencies approximately as~\cite{migdal,eli1st,factor2}
\beg
\omega_q\approx {\omega_0(q)} \sqrt{1-2\lam_0},
\label{renormomega}
\en
where $\lam_0$ is the \textit{bare} electron-phonon coupling constant defined   by the same \eref{lam} as $\lam$, but with $\Omega\to\Omega_0$ and $\omega_q\to\omega_0(q)$.   It follows from \eref{lam} that the \textit{renormalized} dimensionless electron-phonon coupling  is
\beg
\lam =\frac{\lam_0}{1-2\lam_0}.
\label{renlam}
\en
\esref{renormomega} and \re{renlam} are one-loop renormalization equations. They hold for both Holstein~\re{holsteinH} and the more general Hamiltonian~\re{frol1} in 2D and 3D~\cite{agd,dolgov}. In infinite dimensional space~\cite{maksimov},   $1-2\lam_0$ is replaced with $1-\frac{8}{3}\lam_0$ in \esref{renormomega} and~\re{renlam}. 

More accurate renormalization equations are available, but they do not change the fact that the lattice loses stability at a certain $\lam_0=\lam_{\mathrm{LI}}$ and that the renormalized coupling $\lam$ grows monotonously with $\lam_0$ and diverges at $\lam_0=\lam_{\mathrm{LI}}$. Our analysis does not depend on the value of $\lam_\mathrm{LI}$ and for concreteness  we take $\lam_\mathrm{LI}=0.5$. As mentioned in  Introduction, this lattice instability is merely an artifact of the conventional models. It is nevertheless very real in studies   of such models that do not  take precautions to factor it out as we did in this paper.

Main assumptions of the Migdal-Eliashberg theory are that the electron-phonon system is metallic and translationally invariant. None of these are guaranteed past the lattice instability, which changes lattice symmetry and may, for example, open a gap at the Fermi surface through the Peierls mechanism, which we discuss in Sec.~\ref{adiabatic}. It is for this reason that Migdal and Eliashberg restricted~\cite{migdal,eli1st,factor2} their theory to $\lam_0\le 0.5$. Note also that an early textbook account of this theory~\cite{agd}, which closely follows the original work,
makes it clear on p.~182  that $\eta$ (our $\lam_0$) should not be ``too close to $\frac{1}{2}$''.

Equation~\re{renlam} shows that the \textit{entire domain} of the  theory, $0\le\lam\le\infty$, maps to the interval $0\le \lam_0\le 0.5$. 
Asserting its breakdown  past the lattice instability is tautological as such values of $\lam_0$ are already outside of its domain of applicability. A meaningful statement would be that it breaks down at a finite $\lam$, which then maps to a certain $\lam_0 < 0.5$, see also Ref.~\cite{brauer}. Prior work mixes up the true breakdown of the Migdal-Eliashberg theory
with the lattice instability. As a result, it does not  eliminate the possibility that the theory remains valid for all $\lam$, including $\lam=\infty$. This, for example, leaves the door open to the hypothesis~\cite{andrey_validity} that the strong coupling, $\lam\to\infty$, limit of the Migdal-Eliashberg theory  is realized in the Holstein model when $\lam_0\to 0.5$ underscoring  the luck of conclusiveness  of the prior  work. In contrast, our study eliminates this hypothesis.

The confusion stems in part from  misunderstanding of  Migdal's theorem. This theorem is often interpreted as follows: the Migdal-Eliashberg
theory is valid provided the parameter $\lam_0\Omega_0/\eps_F$ is small (for dispersing phonons, we replace $\Omega_0$ with the maximum phonon frequency). This statement is incorrect. This form of the Migdal parameter assumes $T=0$ and $\lam_0$ not too close to 0.5 (no substantial renormalization, i.e., $\lam\sim\lam_0$). The proper zero temperature Migdal parameter, suitable for all $\lam$, uses renormalized coupling and phonon frequency
\beg
x_M=\frac{\lam\Omega}{\eps_F}.
\label{mparameter1}
\en 
Most importantly, Migdal's theorem is a \textit{local} statement about the  Eliashberg stationary point~\cite{meaningmigdal}. It says that \textit{quadratic} fluctuations of the Eliashberg fields $\Sigma_\sigma$ and $\Phi$ around this point are small. This   makes the stationary phase approximation -- the Migdal-Eliashberg theory -- accurate \textit{when} it is the global minimum of the free energy. But it is meaningless to apply  Migdal's theorem as well as the Migdal-Eliashberg theory  when the global minimum is something else, e.g., an insulator, see Fig.~\ref{shape}. 

Moreover, it is not even clear how to evaluate $x_M$ at the ``wrong'' minimum and what is  its significance there. For example, what are the renormalized coupling and phonon frequencies past the lattice instability? The Fermi energy $\eps_F$ plays a different role in an insulator compared to a metal.
At the same time, Migdal's theorem as formulated above \textit{remains valid} when applied at the Eliashberg stationary point even when this point is no longer the global minimum. However, we have to keep in mind that now this stationary point is  not relevant to the equilibrium physics.

Consider, for instance, an impressive Monte Carlo study of the square-lattice Holstein model~\cite{scalapino}. This study reports that the deviation of the $s$-wave pair susceptibility from  its Eliashberg value grows  from roughly 1\% to 25\% as $\lam_0$ increases from 0.4 to 0.5.  A part of this deviation must be due   to the Migdal parameter~\re{mparameter1} being finite. Not only is $x_M$ nonzero, but it also diverges  as $(1-2\lam_0)^{-1/2}$ as we approach  $\lam_0=0.5$, though finite $T$ cuts off this divergence~\cite{meaningmigdal}. 
Another contribution is the finite size effect, which turns the sharp   transition  at $\lam_0=0.5$ present in the Holstein model into a crossover over a certain interval of $\lam_0$ around 0.5. Without knowing the magnitude of these contributions to the deviation, it is impossible to tell whether or not it indicates true breakdown of the  theory. 

Our analysis  is very different from previous work. We showed that the Migdal-Eliashberg theory  breaks down at a finite value of the electron-phonon coupling $\lam$ independently of the  underlying microscopic electron-phonon model. We based this conclusion on an unambiguous marker of the breakdown -- negative specific heat. Our value $\lam_*\approx 3.69$ where the specific heat becomes negative translates into $\lam_0\approx 0.44$ according to \eref{renlam}. This appears close to $\lam_0\approx0.4$ reported  in Ref.~\cite{scalapino} as the point where the determinant Monte Carlo computation starts to deviate from the Migdal-Eliashberg prediction. However, it is important to keep in mind that the entire strong coupling regime of the Eliashberg theory maps to the left vicinity of $\lam_0= 0.5$. Because of this and without knowing the systematic error on the number 0.4 it is difficult to draw any conclusion from its proximity to our result.

\section{Classical phonons}
\label{static_sect}

We saw that static deformations $\bar x_\ii$ of the lattice (zero Matsubara frequency phonons)  facilitate  the breakdown of the Migdal-Eliashberg theory. 
They provide a statistically distributed single-particle potential $V_\ii=\alpha\bar x_\ii$ for the electrons, where $\alpha\propto\sqrt{\lam}$.
To understand this better,  consider the strong coupling limit, $\lam\to\infty$, of this theory. Recall the definition of the electron-phonon coupling  $\lam$ for the Holstein model
\beg 
\lam=\frac{g^2}{\Omega^2}=\frac{\nu_0 \alpha^2}{K}.
\en
We see that the strong coupling limit is  the \textit{free ion limit} -- the limit where the spring constant $K$ of  lattice oscillators vanishes. 
 As $K$ decreases, nonuniform thermal averages of $\bar x_\ii$ come at lower and lower elastic   energy cost, while the strength of the potential $V_\ii$ keeps increasing. Inevitably, at a certain point it becomes energetically favorable to generate  a nonuniform average potential $V_\ii$ for the electrons.  
 
Euclidian  Lagrangian corresponding to the Holstein Hamiltonian~\re{holsteinH}   is
\beg
\begin{split}
L=  \sum_{\bm i \bm j, \sigma}  c^*_{\bm i\sigma} G_{0\ii\jj}^{-1} c_{\bm j \sigma}+
& \sum_{\bm i}\left[ \frac{K x^2_{\bm i} }{2} +
\frac{M (\partial_\tau x_{\bm i} )^2}{2  } \right]
\\
& +\alpha \sum_{\bm i \sigma} c^*_{\bm i\sigma} c_{\bm i \sigma}x_{\bm i}.
\end{split}
\label{holL}
\en
The fields $c^*_{\bm i\sigma}$, $c_{\bm i \sigma}$, and $x_\ii$ depend on the imaginary time $\tau$,
$G_{0\ii\jj}^{-1}=\partial_\tau \delta_{\ii\jj}+t_{\bm i\bm j}-\mu\delta_{\ii\jj}$,
and we replaced the arbitrary single-particle Hamiltonian $h_{\ii\jj}$ with a translationally invariant hopping matrix $t_{\ii\jj}$ and the bare
spring constant $K_0$ with the renormalized constant $K$. The  action in the Matsubara frequency representation reads as
\beg
\begin{split}
S= \!\! \sum_{\bm i \bm j,n \sigma}\!\!  c^*_{\bm i\sigma}(n) G_{0\ii\jj}^{-1} c_{\bm j \sigma}(n)+
 \frac{M}{2} \! \sum_{\bm i}\!\left[ \Omega^2+
 \omega_l^2 \right]  x^2_{\bm i}(l)
\\
 +\alpha \sum_{\bm i \sigma} c^*_{\bm i\sigma}(n+l) c_{\bm i \sigma}(n) x_{\bm i}(l),
\end{split}
\label{holaction}
\en
where now
\beg
G_{0\ii\jj}^{-1}=-i\omega_n \delta_{\ii\jj}+t_{\bm i\bm j}-\mu\delta_{\ii\jj},
\label{green-1}
\en
and $n$ and $l$ stand for fermionic and bosonic Matsubara frequencies $\omega_n$ and $\omega_l$, respectively.

Integrating out the phonon field $x_\ii(l)$, we obtain the effective electron-electron interaction~\re{Holstein_lambda} for the Holstein model, namely,
\beg
\lambda(\omega_l)= \frac{  g^2}{\omega_l^2+\Omega^2}.
\label{Holstein_lambda1}
\en
In the strong coupling limit, $\Omega=0$ and the interaction blows up at $\omega_l=0$, $\lambda(\omega_l=0)=\lam\to\infty$.  This divergence propagates into the normal self-energy $\Sigma_n$ and gives rise to the divergent imaginary part of the pole of the retarded Green's function~\re{GR}. This $\omega_l=0$ part (zero Matsubara frequency phonons) of the interaction   is responsible for the divergence of the quasiparticle decay rate,
as we already saw above.

This divergence arises from integrating out $\omega_l=0$ phonons, because
this is illegal   in the strong coupling limit. In this limit, $\Omega=0$ and $x^2_\ii(0)$ term is absent from the action~\re{holaction}. The integral over $x_\ii(0)$ is no longer Gaussian and this field therefore cannot be integrated out. Instead,  we incorporate the $\alpha\sum_{\ii\sigma} c^*_{\ii\sigma}(n)c_{\ii\sigma}(n)x_\ii(0)$ term in \eref{holaction} into the single-fermion part by
replacing the hopping $t_{\ii\jj}$ in \eref{green-1} with
\beg
h_{\ii\jj}=t_{\bm i\bm j}+\alpha  \bar x_\ii,
\label{hop+}
\en
where
\beg
 \bar x_\ii\equiv x_\ii(0)= \frac{1}{\beta}\int_0^{\beta}\!\!\!  d\tau\, x_\ii(\tau),\quad \beta=\frac{1}{T}. 
\en
We see again that  static displacements of the ions provide an on-site potential for the electrons.   We also discussed in Sec.~\ref{quasidecay} that variables $ \bar x_\ii$ are classical displacement fields (classical phonons). 

\section{Adiabatic limit}
\label{adiabatic}

To gain further insight into post-Migdal-Eliashberg physics,  consider the adiabatic limit where the ion mass $M\to\infty$. 
This limit is complimentary to the strong coupling limit $K\to0$.  All    phonons are classical in the adiabatic limit and their role   becomes especially transparent. 
Studies of polarons, bipolarons  etc. frequently employ this limit as it is much simpler than dealing with quantum phonons~\cite{millis,alexandrov,scalapino,kabanov}. In this limit, $g=0$ and the electron-electron interaction~\re{Holstein_lambda} vanishes for all $\omega_l\ne 0$.

 The Holstein Hamiltonian~\re{holsteinH} becomes at $M=\infty$
\beg
H=\sum_{\bm i \bm j \sigma} t_{\bm i\bm j} c^\dagger_{\bm i\sigma} c_{\bm j \sigma}+  \sum_{\bm i}   \frac{K_0 x_\ii^2}{2}  +
\alpha \sum_{\bm i} n_{\bm i} x_\ii,
\label{holsteinH1}
\en 
where we replaced the arbitrary $h_{\ii\jj}$ with a translationally invariant hopping matrix $t_{\ii\jj}$.
Note that the dimensionless electron-phonon coupling~\re{lam}
\beg
\lam_0=\frac{\nu_0\alpha^2}{K_0}
\en
 remains finite in this limit~\cite{valence}. We use unrenormalized version of \eref{lam}, because there is no renormalization in the usual sense in the adiabatic limit (see below).
Ion displacements $x_\ii$ now commute with the Hamiltonian, which allows us to treat them as classical variables. However, they do not commute with the total momentum operator $\bm P$ and the commutation relations
 \beg
 [x_\ii, H]=[\bm P, H]=0,\quad[x_\ii, \bm P]\ne 0
 \en
  imply that the eigenstates of the Hamiltonian are degenerate~\cite{degenerate}. 

Consider the Holstein Hamiltonian~\re{holsteinH1}. We are to find a lattice distortion $x_\ii$ that minimizes the energy. Suppose we observe that initially uniform $x_\ii$ (independent of $\ii$) become nonuniform  as we increase $\alpha$. This is known as  Peierls or, more generally,  charge density wave (CDW) instability~\cite{pouget,acta}.    Peierls distortion lowers the energy by opening a gap at the Fermi surface  resulting in a metal-insulator transition. 
  In 1D the CDW wavevector is $2p_F$ -- twice the Fermi momentum. In 2D we expect the CDW wavevectors  to depend on the geometry of the Fermi surface as well. The Fourier transform of  $x_\ii$ can now contain  more than one  Fourier mode unlike in 1D. 

CDW instability in dimensions higher than one is a more complicated matter. 1D Fermi surface is perfectly nested at $2p_F$. The closest 2D analog   in the Holstein model~\re{holsteinH1} is a square lattice at half-filling with nearest neighbor hopping. Then, the Fermi surface is a square nested at   $\bm Q=(\pi, \pi)$ and we expect this to be the dominant CDW wavevector. Commensurate $(\pi, \pi)$ CDW has been found in a very similar model at 0.4 filling~\cite{scalapino}, but it could be difficult to differentiate numerically between $(\pi, \pi)$ and nearby wavevectors on a small lattice. And in any case  there is no  reason to expect pure commensurate $(\pi, \pi)$ CDW away  from half filling. Even at half filling there is an admixture of other wavevectors in the CDW~\cite{complicated}.

Nevertheless, let us take the $(\pi, \pi)$ lattice distortion pattern
\beg
x_\ii=X_\mathrm{c.m.} + (-1)^{i_x+i_y} \delta x.
\en
as our variational wavefunction. 
The center of mass displacement $X_\mathrm{c.m.}$ couples only to the total fermion number.  At the minimum $X_\mathrm{c.m.}=-\alpha/K_0$. The Hamiltonian for the remaining degrees of freedom in the momentum representation is
\beg
\begin{split}
H=&\frac{1}{2}\sum_{\bm k\sigma} \left[ \eps_\kk c^\dagger_{\kk\sigma} c_{\kk\sigma} + \eps_{\kk+\bm Q} c^\dagger_{\kk+\bm Q,\sigma} c_{\kk+\bm Q,\sigma}+\right.\\
&\left.\Delta_P c^\dagger_{\kk\sigma} c_{\kk+\bm Q,\sigma} +\Delta_P c^\dagger_{\kk+\bm Q,\sigma} c_{\kk \sigma}   \right] +\frac{\nu_0 N \Delta_P^2}{2\lam_0},
  \end{split}
\en
where $\Delta_P=\alpha\delta x$ is the Peierls gap. 

It is straightforward to diagonalize this Hamiltonian by a Bogoliubov transformation,
\beg
H=\frac{1}{2}\sum_{\bm k\sigma} E_\kk( a^\dagger_{\kk \sigma +}  a_{\kk \sigma +} -  a^\dagger_{\kk \sigma -}  a_{\kk \sigma -} )+\frac{\nu_0 N \Delta_P^2}{2\lam_0},
\label{bcslike}
\en
 where $(a^\dagger_{\kk \sigma \pm} , a_{\kk \sigma \pm})$ are the new quasiparticles and $E_\kk=\sqrt{\eps_\kk^2+\Delta_P^2}$. 
 The Hamiltonian~\re{bcslike} is nearly identical to the mean-field BCS Hamiltonian. Minimizing the total energy with respect to $\Delta_P$, we obtain a version of   the BCS gap equation
 \beg
 \int_{\eps_<}^{\eps_F} \frac{\Delta_P d\eps}{ \sqrt{\eps^2 +\Delta_P^2}}=\frac{\Delta_P}{\lam_0},\quad \eps_<=\eps_F|1-2\mathsf{f}|, 
 \en
 where $\mathsf{f}$ is the filling fraction. For simplicity, we took the density of states to be constant as its energy dependence is unimportant for our discussion. 
 
 As usual, $\Delta_P=0$ is always a solution of the gap equation. A nonzero solution, when it exists, is always the minimum of the energy. 
 At half filling $\eps_<=0$ and the Peierls gap $\Delta_P=\eps_F e^{-1/\lam_0}$ opens already at $\lam_0=0^+$.  Away from the half-filling,  the gap opens at $\lam_0^c = -\left(\ln|1-2\mathsf{f}|\right)^{-1}$ for  this lattice distortion pattern~\cite{var}. The transition is always second order, even though numerically it is easy to mistake it for the first order transition~\cite{scalapino} due to a rapid rise of $\Delta_P$ past $\lam_0^c$ for certain choices of  parameters.
 
 In this example, the metal-insulator transition occurs  at $\lam_0^c=0^+,$ because  the conditions for it are ideal: frozen lattice vibrations and nested Fermi surface. The Migdal-Eliashberg theory applies only at $\lam_0=0$ in this setup. In other circumstances, the transition shifts to nonzero $\lam_0$. In the adiabatic limit, the phonon mediated electron-electron interaction is extremely retarded, $\lam_{nm}=\lam\delta_{nm}$. \eref{gapeq} then implies that the gap function vanishes. The electron Green's function is given by \eref{GR} now for all $T$, because $T_c=0$. Notice that it reproduces  exact energy levels of the Hamiltonian~\re{holsteinH1}  for $\lam_0 <\lam_0^c$ when $x_\ii=\mbox{const}$. Therefore  the Migdal-Eliashberg theory is exact at $T=0$ in the metallic phase, though this phase is confined to $\lam_0=0$. 
 
 Renormalization equations discussed in Introduction do not work in the adiabatic limit, since $\Omega_0=0$ and the phonon propagator vanishes at all but zero frequency.  Temperature-dependent renormalization of the spring constant $K_0$ and $\lam_0$ with it can occur, but we do not investigate it here.
 Most importantly,   this example confirms once more  that the role of the classical phonons is to modify the single-fermion spectrum.

\section{Lattice-Fermionic Superfluidity}
\label{new_theory_sect}

We saw  that classical (zero Matsubara frequency) phonon field $ x_\ii(0)$ provides a statistically distributed potential $V_\ii=\alpha  x_\ii(0)$ for the electrons. As $\alpha$ grows, $x_\ii(0)$ acquire nonzero thermal averages. The resulting single-particle potential  $V_\ii$ together with dressing of fermions by quantum phonons lead to   abrupt changes in the fermion band structure. Examples include gap opening at the Fermi level resulting in a superconductor-insulator transition  and  polaronic Fermi liquid at low densities. At even stronger electron-phonon, interaction dramatic band narrowing and Bose-Einstein condensation of bipolorons~\cite{alexandrov} can occur.

In this section, we   construct a theory  which treats the classical phonons properly. At not too large electron-phonon coupling $\lam$, it reduces to the Migdal-Eliashberg theory, and in the adiabatic limit it reduces to the polaron formation theory, which predicts  electron localization in 2 and 3D for $\lam_0\gtrsim 1$~\cite{kabanov}.  It also reproduces the results of the previous section for the half-filled Holstein model on square lattice in the adiabatic limit.  It continues to work past $\lam_c$ where the Eliashberg theory breaks down and describes at least some of the new phases that emerge at $\lam>\lam_c$.  We dub this theory  lattice-fermionic superfluidity, because it potentially encompasses 
several superfluid phases and because  the lattice (quantum and classical phonons) and the fermions are much closer intertwined in this theory than in the theory of conventional superconductivity.  However, we stress that our theory is meant to describe nonsuperfluid phases, such as a metal or an insulator, as well.

Our starting point is the action~\re{holaction} for the Holstein model where $t_{\ii\jj}$ has been replaced with $h_{\ii\jj}$ given
by \eref{hop+}. In the previous paper~\cite{spinchain}, we determined the effective action and a \textit{spatially nonuniform} version of the Eliashberg stationary point for the Holstein model with an \textit{arbitrary} single-electron Hamiltonian $h_{\ii\jj}$ (see Appendix~A~3 of Ref.~\cite{spinchain}) and used  it to map the free energy to a classical spin chain. The approach is similar to the one outlined  in Sec.~\ref{free_en_sub}, except we now do not assume translational invariance and work in the eigenbasis of an arbitrary $h_{\ii\jj}$.  

The derivation of the theory of lattice-fermionic   superfluidity goes through the same steps except: (a) the effective electron-electron interaction now excludes $\omega_l=0$, because we do not integrate out $\bar x_\ii\equiv x_\ii(0)$, and (b) we need to minimize with respect to the new  parameters $\bar x_\ii$. 
We obtain the following effective action [cf. \eref{feH1}]:
\beg
\begin{split}
{ S}_\mathrm{eff}= T \nu_0 \sum_{nl\alpha}\left[ (\Phi_{n+l}^\alpha)^*  \Lambda_l\Phi_{n}^\alpha + \Sigma_{n+l}^\alpha  \Lambda_l \Sigma_{n}^\alpha \right.\\
\left. - \chi_{n+l}^\alpha  \Lambda_l \chi_{n}^\alpha  \right]
-\sum_{n\alpha} \ln \left[ (\omega_n+\Sigma_{n}^\alpha )^2 +| \Phi_{n}^\alpha |^2\right.\\
\left. + (\chi_{n}^\alpha +\xi_\alpha)^2\right]
+\frac{K_0}{2T}\sum_\ii \bar x_\ii^2.
\end{split}
\label{epseff}
\en
Here $\Phi_n^{\alpha}$, $\Sigma_n^{\alpha}$, and $\chi_n^{\alpha}$ are the components of the three fields $\Phi_\ii(\tau',\tau)$,  $\Sigma_{\ii\up}(\tau',\tau)$, and $\Sigma_{\ii\dn}(\tau',\tau)$ with which we decoupled the four-fermion term after integrating out the phonons. On the stationary point, the fields depend only on the difference $\tau'-\tau$. Let $\Sigma_{\ii n\up}$ be the Fourier transform  of $\Sigma_{\ii\up}(\tau'-\tau)$ with respect to $\tau'-\tau$. We define $\Sigma_{\up n}^{\alpha}$ as
\beg
\Sigma_{\up n}^{\alpha}=\sum_\ii \pi^*_{\ii\alpha} \Sigma_{\ii n\up } \pi_{\ii\alpha},
\en
and similarly for the other fields. Here $\pi_{\ii\alpha}$ are the eigenstates of $h_{\ii\jj}=t_{\ii\jj} +\alpha  \bar x_\ii\delta_{\ii\jj}$, i.e.,
\beg
\sum_\jj \left[ t_{\ii\jj} +\alpha  \bar x_\ii\delta_{\ii\jj} \right] \pi_{\jj\gamma}  = \eps_\gamma \pi_{\ii \gamma}
\label{self-Shrod}
 \en
 and $\xi_\gamma=\eps_\gamma-\mu$. The fields $\Sigma_{n}^\alpha$ and $\chi_{n}^\alpha$ are defined through
 \beg
\Sigma_{n}^\alpha =\frac{ \Sigma_{\up n}^\alpha  - \Sigma_{\dn, -n}^\alpha }{2},\quad i \chi_{n}^\alpha =\frac{ \Sigma_{\up n}^\alpha +\Sigma_{\dn, -n}^\alpha  }{2}.
\label{oddevennu1}
\en
  We retain unrenormalized spring constant $K_0$ for the classical phonons.   
  
 We need to minimize the effective action~\re{epseff} with respect to the real fields $\bar x_\ii$, $\Sigma_{\ii n}$, and $\chi_{\ii n}$ and complex field $\Phi_{\ii n}$.  Minimizing with respect to the latter three fields, we obtain three generalized Eliashberg equations~\cite{spinchain}:
 \beg
\begin{split}
\sum_\gamma \Phi_n^\gamma |\pi_{\ii\gamma}|^2 &=  T \sum_{m\ne n,\gamma}  \frac{\lam_{nm}}{\nu_0}  \frac{\Phi_m^\gamma |\pi_{\ii\gamma}|^2}{\Theta_m^\gamma },\\
\sum_\gamma \Sigma_n^\gamma |\pi_{\ii\gamma}|^2 & =   T \sum_{m\ne n,\gamma}  \frac{\lam_{nm}}{\nu_0} \frac{(\omega_m+\Sigma_m^\gamma) |\pi_{\ii\gamma}|^2}{ \Theta_m^\gamma},\\
\sum_\gamma \chi_n^\gamma |\pi_{\ii\gamma}|^2 & = -  T \sum_{m\ne n,\gamma}  \frac{\lam_{nm}}{\nu_0} \frac{(\xi_\gamma+\chi_m^\gamma) |\pi_{\ii\gamma}|^2 }{ \Theta_m^\gamma},
\end{split}
\label{elialleqnu}
\en
where $\Theta_m^\gamma=(\omega_m+\Sigma_{m}^\gamma)^2 +| \Phi_{m}^\gamma|^2 + (\chi_{m}^\gamma +\xi_\gamma)^2$,
$\lam_{nm}=\lam(\omega_n-\omega_m)$ is given by \eref{Holstein_lambda} as before. The renormalized frequency $\Omega$ in \eref{Holstein_lambda} is an \textit{independent parameter} not fixed by the theory.

 To minimize \eref{epseff} with respect to  $\bar x_\ii$, we use
\beg
\frac{\partial\xi_\gamma}{\partial  \bar x_\ii}= \frac{\partial\eps_\gamma}{\partial \bar x_\ii}=\alpha | \pi_{\ii  \gamma}|^2,
\en
which follows from the first order of the perturbation theory in $\delta\bar x_\ii$. Note that the chemical potential $\mu$ in $\xi_\gamma=\eps_\gamma-\mu$ is a Lagrange  multiplier that does not depend on $\bar x_\ii$ until later, when we fix the average electron number. Setting the derivative of the effective action with respect to $\bar x_\ii$ to zero, we find
\beg
K_0\bar x_\ii  =  -2\alpha\sum_\gamma n_\gamma | \pi_{\ii \gamma}|^2,
\label{fi0}
\en
where $n_\gamma$ are the occupation numbers
\beg
 n_\gamma= \frac{1}{2} - T \sum_m \frac{\xi_\gamma }{(\omega_m+\Sigma_{m}^\gamma )^2 +| \Phi_{m}^\gamma |^2 + \xi_\gamma^2}.
\label{number}
\en
We derived   this expression with the help of the normal Green's function~\re{gsym} generalized   to the case of nonuniform self-energies (replace  $\pp$ with $\gamma$ and $\Sigma_n$ and $\Phi_n$ with $\Sigma_n^\gamma$ and $\Phi_n^\gamma$). The average electron number $N_e=\sum_\gamma 2 n_\gamma$ determines the chemical potential.

Substituting  $\bar x_\ii$ from \eref{fi0} back into \eref{self-Shrod}, we arrive at a discrete nonlinear Schr\"odinger equation where the potential is a weighted  sum of $|\pi_{\ii\gamma}|^2$ over all states,
\beg
 \sum_\jj t_{\ii\jj} \pi_{\jj\gamma} - 4E_b \pi_{\ii\gamma} \sum_\delta n_\delta |\pi_{\ii\delta}|^2=\eps_\gamma\pi_{\ii\gamma}.
 \label{nse}
 \en
 Here $E_b=\frac{\alpha^2}{2 M\Omega_0^2}=\frac{\lam_0}{2\nu_0}$ has the meaning of the polaron binding energy~\cite{suzuki}.
Together with \esref{elialleqnu}  we have four  coupled equations for four unknowns: $\pi_{\ii\gamma}$, $\eps_\gamma$, $\Phi_n^\gamma$, $\Sigma_n^\gamma$, and $\chi_n^\gamma$. 

These equations have several kinds of solutions.
First, there is always the  solution where $\pi_{\ii\gamma}$ are plane waves. In this case, $\alpha x_\ii=\bar\mu$ is spatially uniform and reduces to a shift of the chemical potential, $\mu\to\mu-\bar\mu$ in \esref{elialleqnu}. And conversely,  if $\bar x_\ii$ does not break the translational symmetry of the lattice, i.e., is $\ii$-independent, $\pi_{\ii\gamma}$ are plane waves.  Then, the fields $\Phi_n^\gamma\equiv \Phi_n$,
$\Sigma_n^\gamma\equiv \Sigma_n$, and $\chi_n^\gamma\equiv \chi_n$    are independent of the index $\gamma$ and summing over  it, we  end up with the Eliashberg equations generalized to the non-particle-hole-symmetric case [Eq.~(A.18) in Ref.~\cite{spinchain}], except   $m=n$ terms are absent from the summations.  But as we mentioned above, this is an an alternative  way to write the Eliashberg equations. Indeed, we showed in Ref.~\cite{spinchain} that \esref{elifamiliar3} and \esref{elifamiliar311} are equivalent. The same applies to the more general   Eliashberg equations for the fields $\Sigma_n$, $\Phi_n$, and $\chi_n$.

Now consider the adiabatic limit. In this limit, $\lam_{nm}=0$ for $m\ne n$ and only the nonlinear Schr\"odinger equation~\re{nse} is left.  This equation  describes polarons in 1, 2, and 3D, see Ref.~\cite{kabanov} and references therein. Setting additionally $T=0$, we see that \eref{nse} is the exact minimization condition for the Holstein Hamiltonian~\re{holsteinH1} from which we deduced that the system becomes a CDW insulator for $\lam_0>\lam_0^c$.  Therefore the  lattice-fermionic theory remains valid long after the Eliashberg theory breaks down and is exact  in the adiabatic limit for any value of $\lam_0$, at least at $T=0$.

As the strength of the electron-phonon interaction $\alpha$ grows, the potential in \eref{nse} becomes stronger. The electron effective mass grows and the band narrows. The band narrowing is  exponential in $-E_b/\Omega_0$~\cite{holstein,alex2}. In the narrow band regime, \eref{nse} supports self-trapping of fermions (polarons). Indeed, consider the flat band limit for simplicity. Let $\pi_{\ii\gamma}=\delta_{\ii\bm\gamma}$ be a state where the fermion is at site $\bm\gamma$. We see that by occupying certain   sites, the fermions make the potential~\re{fi0} deeper at these sites thus lowering their energy.  In this regime, solutions of \esref{nse} and \re{elialleqnu} are  well outside of the Migdal-Eliashberg theory.   The system of equations \re{nse} and \re{elialleqnu} is more complex than the Eliashberg gap equation~\re{gapeq}. Nevertheless, it is still solvable in a polynomial time as the number of equations and unknowns is  polynomial in the number of sites and Matsubara frequencies kept in the simulation.

The accuracy of the lattice-fermionic theory  in the regime where the quasiparticle bandwidth is no longer the largest energy scale requires further investigation, but, at least at the first glance, it appears to have the potential to describe many different phases, such as the polaronic metal and polaronic BCS  condensate. It is interesting to understand how our theory compares to the traditional approaches to these phenomena, e.g., to those based on the Holstein-Lang-Firsov transformation~\cite{holstein,alex2}.

\section{Summary and Outlook}

We showed in this paper that the Migdal-Eliashberg theory breaks down when the actual electron-phonon coupling $\lam$ exceeds $\lam_c$, where $3.0\lesssim \lam_c \lesssim 3.7$. The breakdown is marked by negative quasiparticle  heat capacity   of the Migdal-Eliashberg normal state at $\lam> 3.69$ in a range of temperatures above the superconducting transition temperature. Another pathology is the quasiparticle decay rate $\Gamma=\pi\lam T \gg T$ at strong coupling.
These findings indicate that the electron-phonon system cannot be in the state prescribed by this  theory as  it is thermodynamically unstable. A new phase therefore must emerge for $\lam>\lam_c$ below a certain critical temperature $T_{c1}$.   

The new phase breaks the translational invariance of the crystal because strong electron-ion Coulomb interaction, $\lam>\lam_c$, is incompatible with uniform electron charge distribution. Instead, a lattice distortion similar to the Peierls transition occurs at $\lam_c$ that brings electrons on average closer to the ions.  More precisely, this is   a ``many-body Peierls transition'' as the electron-electron interactions mediated by quantum phonons play a critical role in it. This transition is marked by an abrupt change of the quasiparticle spectrum near the Fermi level.

We saw in our previous work~\cite{spinchain}   that solutions of  Eliashberg equations correspond to stationary points of the free energy functional.  The superconducting stationary point continues to exist for $\lam>\lam_c$ below the critical temperature $T_{c2}$, though it is no longer the global minimum of the free energy. We showed above that
$T_{c2}<T_{c1}$ and that this implies a first order phase transition as a function of $\lam$ between the Migdal-Eliashberg superconducting state and the new phase. 
Depending on the  filling fraction, crystal symmetry  and other parameters, the new phase
can be a CDW insulator or a Fermi liquid with broken lattice translational symmetry.  

We proposed a new theory --  lattice-fermionic theory of superfluidity -- that bridges the gap between the Migdal-Eliashberg theory and   phases that emerge at stronger coupling.
The idea is to incorporate the static distortion of the lattice into the single-particle Hamiltonian as a variable potential for the fermions. We treat the phonon mediated electron-electron interactions in a manner similar to the Migdal-Eliashberg theory. However, now the self-energy fields $\Phi_n^\alpha$, $\Sigma_n^\alpha$, and $\chi_n^\alpha$ depend on  single-particle states $|\alpha\rangle$. The theory does not assume translational invariance. We derived the effective action for these fields and  lattice distortions and determined its stationary point. The outcome is a set of four coupled equations. Three of them are equations
for the self-energies. The fourth equation is a nonlinear Schr\"odinger equation for the single-particle spectrum.  At small $\lam$ our theory reproduces the Migdal-Eliashberg theory. Past $\lam_c$ it captures the insulating phase and at least some of the polaron physics.

An apparent open problem is to investigate the phase diagram of the lattice-fermionic theory at strong coupling and to compare it to existing studies of  many-body electron-phonon physics beyond the Migdal-Eliashberg theory. Even though the equations we derived are significantly more complicated than Eliashberg equations in their simplest form, we believe our theory is nevertheless  quite amenable to both
computational and analytic treatments.  

Note  that our study implies an upper bound on the ratio of the critical temperature $T_c$ to the characteristic phonon frequency for conventional superconductors. We use the strong coupling asymptote $T_c\approx 0.183 \sqrt{\lam}\omega_\mathrm{ln}$. Here $\omega_\mathrm{ln}$ is the characteristic bosonic frequency defined through $\ln\omega_\mathrm{ln}=\langle\ln\omega\rangle,$ where $\langle\ln\omega\rangle$ is the spectral average of the log of the bosonic frequency.
This formula fits $T_c$ of superconductors with $\lam\ge 2.25$ quoted in Ref.~\cite{carbotte} reasonably well. We established above that $\lam_c\le 3.69$.  It follows that $T_c/\omega_\mathrm{ln} \le 0.35,$ cf.  upper bound proposed in Ref.~\cite{bound}.

\begin{acknowledgments}

We thank  I. L. Aleiner, A. V. Chubukov, and I. V. Lerner for helpful discussions.

\end{acknowledgments}

\onecolumngrid 

\appendix 
\section{Low temperature entropy and specific heat   in $\lambda\to\infty$ limit}
\label{matsubara}

In this Appendix, we  outline the calculation of the entropy and specific heat in the superconducting state at low
temperatures for $\lam=\infty$ ($\Omega=0$).  
 In the main text we derived \eref{mats_sum} for the free energy difference $\delta f=f_s-f_n$ between superconducting and normal states
 \beg
 \frac{d  [\delta f]}{ dT}= -8\pi\nu_0 \sum_{n=0}^\infty \left(\frac{\omega_n^2}{\sqrt{\omega_n^2+\Delta_n^2}}-\omega_n\right).
 \label{mats_sum2}
 \en
 It remains to evaluate the sum over the Matsubara frequencies. We do so with the help of the Poisson summation formula~\cite{grosso}
 \beg
 \begin{split}
 \sum_{n=0}^\infty \mathsf{h}(\omega_n)=\frac{1}{T} \int_0^\infty \mathsf{h}(\omega) \frac{d\omega}{2\pi}+\frac{\pi T}{12}\mathsf{h}'(0)
-\sum_{s=1}^\infty \frac{ 2 (-1)^s T}{ s^2} \int_{0}^\infty \mathsf{h}''(\omega) \cos\left(\frac{\omega s}{T}\right) \frac{d\omega}{2\pi}.
\end{split}
\label{possion_sum}
 \en
 At low $T$ it is sufficient to keep only the $s=1$ term in the summation over $s$ as other terms are exponentially smaller. In our case
 \beg
 \mathsf{h}(\omega)=\frac{\omega^2}{\sqrt{\omega^2+\Delta^2(i\omega)}}-\omega.
 \en
 We need the following two integrals:
 \beg
  I_1=\int_0^\infty\!\!\! d\omega \left(\omega-\frac{\omega^2}{\sqrt{\omega^2+{\Delta }^2(i\omega)}}\right), \quad
  I_2= - T^2 \int_0^\infty\!\!\! d\omega \left(\frac{\omega^2}{\sqrt{\omega^2+{\Delta }^2(i\omega)}}\right)''
  \cos\left(\frac{\omega }{T}\right),
  \label{integrals}
 \en
 where $\Delta(i\omega)$ is the solution of the $T=0, \Omega=0$ version of the gap equation~\re{gapeqn}
 \beg
\omega\sin\theta =   \frac{g^2}{2} \int_{-\infty}^\infty d\tilde\omega \frac{\sin(\tilde \theta -\theta)}{(\omega-\tilde\omega)^2 },\quad \theta\equiv  \theta(\omega),\quad  \tilde \theta\equiv  \theta(\tilde \omega)
\label{gapeqn1}
\en
Equations~\re{mats_sum2} and \re{possion_sum} then imply
\beg
 \frac{\phantom{.}d    f_s}{ dT}= \frac{\phantom{.} d    f_n}{ dT}+\frac{4\nu_0  I_1}{T}+\frac{2\pi^2\nu_0 T}{3}+\frac{8\nu_0  I_2}{T}.
 \label{ss}
 \en
 Interestingly, we are able to obtain an exact answer for $I_1$, $4I_1= g^2$, which can be interpreted as a sum rule that the zero temperature gap function on the Matsubara axis must satisfy in the strong coupling limit.
 
 Recall that
 \beg
 \cos\theta(\omega)=\frac{\omega}{\sqrt{\omega^2+\Delta^2(i\omega)}},\quad  \sin\theta(\omega)=\frac{\Delta(i\omega)}{\sqrt{\omega^2+\Delta^2(i\omega)}}.
 \label{theta_delta1}
 \en
In terms of $\theta(\omega)$ the expression for $I_1$ reads as
\beg
I_1=\int_0^\infty d\omega\omega(1-\cos\theta)=- \frac{1}{4} \int_{-\infty}^\infty d\omega \, \omega^2 \theta'  \sin\theta.
\en
Here we integrated by parts taking into account that $\Delta(i\omega)$ is even in omega and $\Delta(i\omega)\to0$ as $\omega\to0$~\cite{spinchain}. The same integral appears if we integrate the gap equation~\re{gapeqn1} over $\omega$ and then perform integrations by parts with respect to $\omega$ on the left hand side and with respect to both $\omega$ and $\tilde\omega$ on the right hand side. We obtain
\beg
I_1=\frac{g^2}{8} \int\limits_{-\infty}^\infty\!\!\! d\omega \!\!\! \int\limits_{-\infty}^\infty \!\!\!  d\tilde\omega \frac{\omega}{\omega-\tilde\omega}
\cos(\theta-\tilde\theta)\theta' \tilde\theta'.
\label{interm123}
\en
 Using $\omega/(\omega-\tilde\omega)=1+ \tilde\omega/(\omega-\tilde\omega)$, we rewrite \eref{interm123} in the form
 \beg
 I_1=\frac{g^2}{8} \int\limits_{-\infty}^\infty\!\!\! d\omega \!\!\! \int\limits_{-\infty}^\infty \!\!\!  d\tilde\omega \cos(\theta-\tilde\theta)\theta' \tilde\theta'+\frac{g^2}{8} \int\limits_{-\infty}^\infty\!\!\! d\omega \!\!\! \int\limits_{-\infty}^\infty \!\!\!  d\tilde\omega \frac{\tilde\omega}{\omega-\tilde\omega}
\cos(\theta-\tilde\theta)\theta' \tilde\theta'.
\label{interm124}
\en
Interchanging $\omega$ and $\tilde\omega$ in the last integral and comparing to \eref{interm123}, we notice that it is equal to $-I_1$. Therefore
 \beg
 2I_1=\frac{g^2}{8} \int\limits_{-\infty}^\infty\!\!\! d\omega \!\!\! \int\limits_{-\infty}^\infty \!\!\! d\tilde\omega \cos(\theta-\tilde\theta)\theta' \tilde\theta' = \frac{g^2}{8} \int\limits_{0}^\pi\!\! d\theta \!\! \int\limits_{0}^\pi \!\! d\tilde\theta \cos(\theta-\tilde\theta)=\frac{g^2}{2}.
 \en
 Going back to the definition of $I_1$ in \eref{integrals}, we see that we derived an identity
 \beg
 \int_0^\infty\!\!\! d\omega \left(\omega-\frac{\omega^2}{\sqrt{\omega^2+{\Delta }^2(i\omega)}}\right)=\frac{g^2}{4},
 \en
 for the Eliashberg gap function $\Delta(i\omega)$ at zero temperature and $\lam=\infty$.
 
 Replacing $4I_1$ with $g^2$ in \eref{ss} and substituting $f_n$ from \eref{nsf_strong}, we obtain the following expression for the entropy of the superconducting state:
 \beg
 S_s= -\frac{\phantom{.}d    f_s}{ dT}=  -\frac{8\nu_0  I_2}{T}.
 \label{ss1}
 \en
 Note that the second and third terms on the right hand side of \eref{ss}  cancel  the entropy of the normal state. To determine $I_2$,
 we first integrate by parts twice casting it into the form
 \beg
 I_2= - T^2 \int_0^\infty\!\!\! d\omega \left(\frac{\omega^2}{\sqrt{\omega^2+{\Delta }^2(i\omega)}}\right)''
  \cos\left(\frac{\omega }{T}\right)=\frac{1}{2}\int_{-\infty}^\infty\!\!\! d\omega  \frac{\omega^2}{\sqrt{\omega^2+{\Delta }^2(i\omega)}} 
   e^{  i\omega/T }.
 \en 
 We turn the last integral   into a contour integral by closing the contour in the upper half plane of complex $\omega$.  We saw in Sec.~\ref{selfsect} that the function $i\omega/\sqrt{\omega^2+{\Delta }^2(i\omega)}$    has simple poles at points where the square root vanishes~\cite{omega}, see also Ref.~\cite{combescot}. These are poles rather than branching points because  roots of   $\omega^2+{\Delta }^2(i\omega)=0$ are doubly degenerate. The poles are at a discrete set of points along the imaginary axis,
 \beg
 \omega=\pm i E_n,\quad n=1,2,3,\dots,
 \en
  where $E_n$ are real and positive and have the meaning of   single fermion energy levels. In particular, $E_1$ is the energy gap. This shows that the density of state is a sum of delta functions centered at $E_n$, i.e.,  the excitation spectrum is discrete in the strong coupling limit~\cite{combescot} provided, of course, this limit is physical in the first place.   
  
 It is now straightforward to evaluate $I_2$ by the residue theorem. The contribution from $E_1$, the pole closest to the real axis, is exponentially larger than that from all other $E_n$. Taking   the residue at the pole and $E_1$ from Ref.~\cite{combescot}, we find that the leading low $T$ asymptotic behaviors of the entropy $S_s$ and the specific heat 
 $C_s=T dS_s/dT$ are
 \beg
 S_s\approx 17.84 \nu_0 \frac{E_1}{T} e^{-E_1/T},\quad C_s\approx 17.84 \nu_0 \left(\frac{E_1}{T}\right)^2 e^{-E_1/T},\quad E_1\approx 1.16g.
 \en 
 
\section{Band structure in the strong coupling regime}
\label{correction_sec}

We saw in the main text that  the quasiparticle spectrum is discrete in the superconducting state at $\lam=\infty$. Here, by analyzing the gap equation on the real frequency axis, we show that at finite $\lam$ the discrete energy levels broaden into narrow energy bands of width $\Omega=g\lam^{-1/2}$. 

In this Appendix only, we choose the energy units so that
\beg
\mbox{$g=1,\quad$ or, equivalently, $\quad\lam\Omega^2=1.$}
\en
The Eliashberg gap equation continued towards the real axis reads as~\cite{cont1,cont2,combescot}
\beg
\omega D(\omega) B(\omega)-A(\omega)=\frac{\pi}{2\Omega}\left\{ \frac{ D(\omega-\Omega)- D(\omega) }{\sqrt{ D^2(\omega-\Omega)-1} }\left[ n_B(\Omega)+n_F(\Omega-\omega)\right] + (\Omega \to -\Omega) \right\},
\en
where $n_B$ and $n_F$ are  Bose and Fermi distribution functions, respectively, and
\begin{align}
A(\omega)=2\pi T \sum_{n=0}^\infty \frac{\Delta_n (\omega_n^2-\omega^2+\Omega^2)}{X_n(\omega) \sqrt{\omega_n^2+\Delta_n^2} },
\quad B(\omega)=1+4\pi T \sum_{n=0}^\infty \frac{ \omega_n^2 }{X_n(\omega) \sqrt{\omega_n^2+\Delta_n^2} },\label{AB}\\
D(\omega)=\frac{\Delta(\omega)}{\omega},\quad X_n(\omega)=(\omega^2+\omega_n^2)^2+2\Omega^2 (\omega_n^2-\omega^2)+\Omega^4.
\end{align}
Taking $T\to0$ limit, we find
\beg
\omega D(\omega) B(\omega)-A(\omega)=\frac{\pi}{2\Omega}\frac{ D(\omega-\Omega)- D(\omega) }{\sqrt{D^2(\omega-\Omega)-1} }.
\label{comb1}
\en
We are interested in the correction to the strong coupling limit $\Omega\to0$. To obtain it, we expand the right hand side of the above equation to the first order in $\Omega$
 \beg
\frac{2}{\pi}\bigl[\omega D  B -A \bigr]=\frac{D'}{\sqrt{D^2-1} }+ \Omega \frac{ D {D'}^2 }{ (D^2-1)^{3/2} } -\frac{\Omega}{2} \frac{D''}{\sqrt{D^2-1} },
\label{comb2}
\en
where $D\equiv D(\omega)$ and $D'\equiv dD/d\omega$. 

Following Combescot who used \eref{comb2} at $\Omega=0$   to analyze the quasiparticle spectrum in the strong coupling limit~\cite{combescot}, we introduce a new variable $\phi(\omega)$ as $D=1/\sin\phi$. 
\eref{comb2} becomes
\beg
\phi'-\frac{\Omega}{2} \left(\phi''+{\phi'}^2 \tan\phi\right)= \frac{2}{\pi}\bigl[\omega   B-A \sin\phi  \bigr].
\label{phiode}
\en
The density of states~\re{dos} in terms of $\phi(\omega)$ is
\beg
\nu(\omega)=\nu_0\, \mbox{Im} \left[\tan\phi(\omega)\right]. 
\label{dos1}
\en
Combescot showed that $\phi(\omega)$ is real to zeroth order in $\Omega$, and consequently the density of states is a sum of delta functions,
\beg
\nu(\omega)=\pi\nu_0 \sum_{k=1}^\infty P_k\left[ \delta(\omega-E_k) + \delta(\omega+E_k) \right],
\en
where $E_k$ are solutions of $\cos\left[\phi(E_k)\right]=0$, or, equivalently, of $\phi(E_k)=\pi (k -1/2)$. Imaginary part of $\nu(\omega)$
comes from  poles of $\tan\phi(\omega)$ at $\omega=\pm E_k-i0^+$.
It is straightforward to show using \eref{AB} that $A(\omega)$ vanishes as $1/\omega^2$ and $B(\omega)\to 1$ as $\omega\to \infty$. At $\Omega=0$ and large $\omega$,  \eref{phiode} takes the form $\phi'=2\omega/\pi$.  Therefore, $\phi(\omega)\approx \omega^2/\pi\equiv\phi_0(\omega)$ which implies the leading large $k$ asymptotic behavior~\re{largek} of $E_k$.

Let us analyze \eref{phiode} at large $\omega$ and small but finite $\Omega$. Corrections to the right hand side due to finite $\Omega$ are suppressed by a factor of $1/\omega^2$. In the zeroth order in $\Omega$, $\phi'\sim\omega$ and $\phi''\sim 1$. Therefore, the term
containing $\phi''$ is negligible and we have
\beg
\phi'-\frac{\Omega}{2}  {\phi'}^2 \tan\phi = \frac{2\omega}{\pi}.
\label{phiode1}
\en
The $\tan\phi$ term is important near $\omega=E_k$ where $\tan\phi$ diverges. Near these points \eref{phiode1} becomes
\beg
y'+ \frac{\Omega}{2}  \frac{ \phantom{a} {y'}^2 }{y} = \frac{2 E_k}{\pi},
\label{phiode2}
\en
where $y=\phi(\omega) -\phi_0(E_k)=\phi(\omega)-E_k^2/\pi$. Solving for $y'$, we find
\beg
y'=-\frac{1}{\Omega}\left( y+\sqrt{y^2+by}\right),\quad b=\frac{8\Omega E_k}{\pi}.
\label{yode}
\en
The plus sign is dictated by the requirement that  for $\Omega\to0$ we recover the zeroth order equation $y'\approx 2E_k/\pi$.
In zeroth order in $\Omega$,  $y= 2\omega E_k/\pi $ is real and the density of states~\re{dos1} is zero except at $\omega=E_k$ In the next order in $\Omega$, $y$ and therefore $\phi(\omega)$ acquire an imaginary part  proportional to $\Omega$. We see this from \eref{yode} --  the square root is imaginary and of the order $\Omega$ for $-b < y <0$. Upon integration over $\omega$, it gives rise to an imaginary part of $y$  of the order $\Omega$. The density of states~\re{dos1} is therefore nonzero in the interval $(\omega_1, \omega_2)$ of $\omega$ for which $y$ falls in between $-b$ and 0. The length of this interval is the bandwidth we are after.

Equation~\re{yode} integrates by variable separation method to
\begin{align}
\frac{b}{u(y)} - \frac{1}{2} \ln u^2(y)= \frac{2\omega}{\Omega}+\mbox{const},\label{uomega}\\
u(y)=b+ 2y+\sqrt{y^2+by}.
\end{align}
To determine $\omega_1$ and $\omega_2$, we set $y=-b$ and $y=0$, respectively, in the left hand side of  \eref{uomega}. We find
$\omega_2-\omega_1=\Omega$. Thus, the discrete level $E_k$ splits into a narrow band of width $\Omega$ similar to how atomic energy levels split into bands when  atoms form a lattice  and atomic orbitals hybridize.


\begin{thebibliography}{99}
\bibitem{grand} We dub $f$  the free energy even though it is, strictly speaking, the grand potential as it is a function of the chemical potential rather than the particle number.
\bibitem{zeta2} A usual way to derive this is to use the Poisson summation formula discarding the divergent ground state energy. It is worthwhile to note that  the zeta function regularization procedure~\cite{kleinert} leads to the same result. Using $\omega_n=2\pi T(n+\frac{1}{2})$, we have $f=-4\pi \nu_0T^2\sum_{n=-\infty}^\infty |n+\frac{1}{2}|$. Further, $\sum_{n=-\infty}^\infty |n+\frac{1}{2}|=2\zeta(-1,\frac{1}{2})=\frac{1}{24}$, where $\zeta(s, a)$ is the Hurwitz zeta function, and \eref{nif} follows.
\bibitem{nozeta} This divergence cannot be removed by zeta function regularization, because $\zeta(s)$ is singular at $s=1$.
\bibitem{omega} Note that $\omega$ in Sec.~\ref{selfsect} differs from $\omega$ in Appendix~\ref{matsubara} and \eref{iden124} by a factor of $i$.
\bibitem{valence} For an insulator $\nu_0$ can be defined, e.g., as the density of states averaged over the valence band.
\bibitem{degenerate} This algebra implies that translations of the set $\{x_\ii\}\to \{x_{\ii+\bm m}\}$ for any $\bm m$ are also eigenstates with the same energy.  The degree of the degeneracy is  the number of inequivalent sets that such translations generate. For example, 
the configuration $x_\ii=X_\mathrm{c.m.} +(-1)^{i_x+i_y} \delta x$ on a square lattice is twofold degenerate.
\bibitem{var}  However, the $(\pi, \pi)$ variational wavefunction is suboptimal  away from half filling, so the true $\lam_0^c$ must be smaller.
\end{thebibliography}
\end{document}